\documentclass[fleqn,usenatbib]{mnras}
\usepackage{newtxtext,newtxmath}

\usepackage[T1]{fontenc}

\DeclareRobustCommand{\VAN}[3]{#2}
\let\VANthebibliography\thebibliography
\def\thebibliography{\DeclareRobustCommand{\VAN}[3]{##3}\VANthebibliography}

\usepackage{graphicx}
\usepackage{amsmath}
\usepackage[LGRgreek]{mathastext}
\usepackage{array}
\usepackage{mathtools}
\usepackage{booktabs}
\usepackage{float}
\usepackage{subcaption}
\usepackage[english]{babel}
\usepackage{hyperref}
\addto\extrasenglish{
    
}

\title[TRAPUM Large Magellanic Cloud survey I]{The TRAPUM Large Magellanic Cloud pulsar survey with MeerKAT I: Survey setup and first seven pulsar discoveries}

\author[V. Prayag et al.]{\parbox{\textwidth}{
V. Prayag,$^{1,2}$\thanks{E-mail: \href{mailto:venu.prayag@gmail.com}{venuprayag@gmail.com}}
L. Levin,$^{3}$
M. Geyer, $^{2,1}$
B.~W.~Stappers,$^{3}$
E. Carli,$^{3}$
E.~D.~Barr,$^{4}$
R. P. Breton,$^{3}$
S. Buchner,$^{5}$
M. Burgay,$^{6}$
M. Kramer,$^{4}$
A. Possenti,$^{6}$
V. Venkatraman Krishnan,$^{4}$
C. Venter,$^{7,8}$
J. Behrend,$^{4}$
W. Chen,$^{4}$
D.~M. Horn, $^{5}$
P.~V.~Padmanabh,$^{9,10,4}$
A. Ridolfi$^{6,4}$
}
\\ \\ \\
$^{1}$Department of Astronomy, The University of Cape Town, Private Bag X3, Rondebosch 7701, Cape Town, South Africa\\
$^{2}$High Energy Physics, Cosmology \& Astrophysics Theory (HEPCAT) Group, Department of Mathematics \& Applied Mathematics, \\University of Cape Town, Cape Town 7700, South Africa \\
$^{3}$Jodrell Bank Centre for Astrophysics, Department of Physics and Astronomy, The University of Manchester, Manchester M13 9PL, United Kingdom \\
$^{4}$Max-Planck-Institut f\"{u}r Radioastronomie, Auf H\"{u}gel 69, D-53121 Bonn, Germany \\
$^{5}$South African Radio Astronomy Observatory (SARAO), 2 Fir Street, Black River Park, Observatory, Cape Town, 7925 \\
$^{6}$INAF-Osservatorio Astronomico di Cagliari, via della Scienza 5, 09047 Selargius (CA), Italy \\
$^{7}$Centre for Space Research, North-West University, Private Bag X6001, Potchefstroom 2520, South Africa \\
$^{8}$National Institute for Theoretical and Computational Sciences, South Africa \\
$^{9}$Max-Planck-Institut f\"{u}r Gravitationsphysik (Albert-Einstein-Institut), D-30167 Hannover, Germany\\
$^{10}$Leibniz Universit\"{a}t Hannover, D-30167 Hannover, Germany
}
\date{Accepted XXX. Received YYY; in original form ZZZ}

\pubyear{2024}

\begin{document}
\label{firstpage}
\pagerange{\pageref{firstpage}--\pageref{lastpage}}
\maketitle

\begin{abstract}
The Large Magellanic Cloud (LMC) presents a unique environment for pulsar population studies due to its distinct star formation characteristics and proximity to the Milky Way. As part of the TRAPUM (TRAnsients and PUlsars with MeerKAT) Large Survey Project, we are using the core array of the MeerKAT radio telescope (MeerKAT) to conduct a targeted search of the LMC for radio pulsars at L-band frequencies, 856--1712\,MHz. The excellent sensitivity of MeerKAT, coupled with a 2-hour integration time, makes the survey 3 times more sensitive than previous LMC radio pulsar surveys. We report the results from the initial four survey pointings which has resulted in the discovery of seven new radio pulsars, increasing the LMC radio pulsar population by 30 per cent. The pulse periods of these new pulsars range from 278 to 1690\,ms, and the highest dispersion measure is 254.20\,$pc \, cm^{-3}$. We searched for, but did not find any significant pulsed radio emission in a beam centred on the SN\,1987A remnant, establishing an upper limit of 6.3\,$\upmu Jy$ on its minimum flux density at 1400\,MHz.
\end{abstract}


\section{Introduction}

The Large Magellanic Cloud (LMC) and the Small Magellanic Cloud (SMC) are two irregular galaxies and the closest galactic companions to the Milky Way. As a result, the Magellanic Clouds have been an enduring reference for drawing parallels and contrasts with our Galaxy. They offer an ideal setting for the study of stellar populations and how their environments impact the process of star formation. Situated beyond the boundaries of the disc of the Milky Way at a distance of 49.6\,kpc \citep{Graczyk2020}, the LMC benefits from a particularly advantageous observational angle, notably from the Southern Hemisphere, with a lower line-of-sight dust extinction in comparison to our own Galaxy.

The SMC has a low average metallicity of $[Fe/H]=(-0.97 \pm 0.05)\,dex$ \citep{Choudhury2020}, implying that it will have a distinct stellar environment with less massive stars being formed \citep{Pakmor2022}. In contrast, the LMC has a similar mean stellar metallicity, $[Fe/H] = (-0.42 \pm 0.04)\,dex$ \citep{Choudhury2021}, to that of the thin disc of the Milky Way, $[Fe/H] = -0.5\,dex$ \citep{Matteucci2014}, which allows us to compare the star formation histories between the two galaxies. The LMC has a greater abundance of supernova remnants (SNRs, \citealt{Badenes2010, Zangrandi2024}) and high-mass X-ray binaries (HMXBs, \citealt{Liu2005, Antoniou2016}) per unit mass than our Galaxy, giving an indication of the high star formation rate prevailing there \citep{Grimm2003}. Consequently, it is anticipated that a larger proportionate population of isolated young neutron stars (hereafter NSs) and NSs in binary systems, especially with massive star companions, will be found within the LMC. Young pulsars, generally rotation-powered, are often associated with SNRs, and they tend to have irregular rotation patterns which are attributed to `glitches' and timing noise (e.g. \citealt{Lyne1999, Haskell2015, Basu2022}). Accretion-powered pulsars, such as the progenitors to millisecond pulsars (MSPs), are found in binaries. The transfer of angular momentum during the accretion phase of binaries leads to MSPs having pulse periods typically shorter than 10\,ms \citep{Alpar1982, Radhakrishnan1982, Papitto2013}.

The present pulsar population of the LMC consists of 25 pulsars, while the SMC is home to 16 pulsars \citep{Manchester2005}\footnote{\href{https://www.atnf.csiro.au/research/pulsar/psrcat/}{https://www.atnf.csiro.au/research/pulsar/psrcat/} PSRCat v2.3.0}. Out of these, 14 from the SMC (see \citealt{Carli2024}) and 24 from the LMC have been detected in the radio domain. Together, these 38 pulsars constitute the extragalactic rotation-powered radio pulsar population, which accounts for only about 1 per cent of the more than 3500 radio pulsars discovered to date.

 Currently, there is only one known extragalactic binary pulsar system, PSR\,J0045$-$7319, located in the SMC \citep{McConnell1991, Kaspi1994}. The potential discovery of binary systems in the LMC, particularly double NS systems, could provide valuable insights into the extragalactic NS population and the rate of extragalactic NS mergers. In addition, recent evidence for the gravitational wave background using pulsar timing arrays \citep{Agazie2023, Antoniadis2023, Reardon2023, Xu2023} highlights the significance of discovering stable extragalactic MSPs. Pulsars located closely together in the sky exhibit a significantly greater timing correlation due to the gravitational wave Earth-term, compared to pulsars with greater spatial separations. So far, there have been no MSPs found in the Magellanic Clouds. 

Pulsars within the LMC provide a unique line-of-sight that passes through the interstellar medium (ISM) of the LMC and that of the Milky Way, as well as the intergalactic medium. As pulsar signals propagate through these regions, they interact with electrons, causing dispersion that follows a quadratic dependence on radio frequency. This is quantified as the dispersion measure (DM), which depends on the total electron column density between Earth and the pulsar. By finding a large number of radio pulsars in different locations of the LMC, we can probe the different regions, enhancing our understanding of their electron density content.

One telescope which has the required sensitivity and optimal geographical location to efficiently observe the LMC is the Parkes 64\,m radio telescope, recently given the indigenous Wiradjuri name "Murriyang" (hereafter Murriyang), located in New South Wales, Australia. Murriyang has played, and continues to play, an instrumental role in the search for pulsars, having discovered 22 out of the 25 known pulsars in the LMC \citep{McCulloch1983, McConnell1991, Crawford2001, Manchester2006, Ridley2013, Hisano2022}.

MeerKAT \citep{Jonas2016, Camilo2018}, located in the Karoo in the Northern Cape province of South Africa at a latitude of $-$30\textdegree{}43\arcmin \citep{Bailes2020}, is a state-of-the-art radio telescope also well placed to observe the LMC. Transients and Pulsars with MeerKAT\footnote{\href{http://www.trapum.org/}{http://www.trapum.org/}} (TRAPUM) is a large survey project with the aim to search for new pulsars, as well as various transient phenomena at radio frequencies \citep{Stappers2016}. One of the objectives of TRAPUM is to survey the Magellanic Clouds and here, we report the results from the first four pointings of the TRAPUM LMC Survey.

Our paper is structured as follows: in \autoref{section: Previous surveys and discoveries}, we review previous surveys of the LMC and examine the current LMC pulsar population. We describe our approach to source selection and the planning of observations in \autoref{section: Source selection}. In \autoref{section: Observations}, we outline the observation setup and the search process. We present the results obtained from the initial four pointings of the survey in \autoref{section: Results}. Finally, in \autoref{section: Discussion}, we conclude on our findings.

\section{Previous surveys and discoveries}\label{section: Previous surveys and discoveries}

\begin{center}
\begin{figure*}
\begin{tabular}{cc}
\includegraphics[width=.46\linewidth]{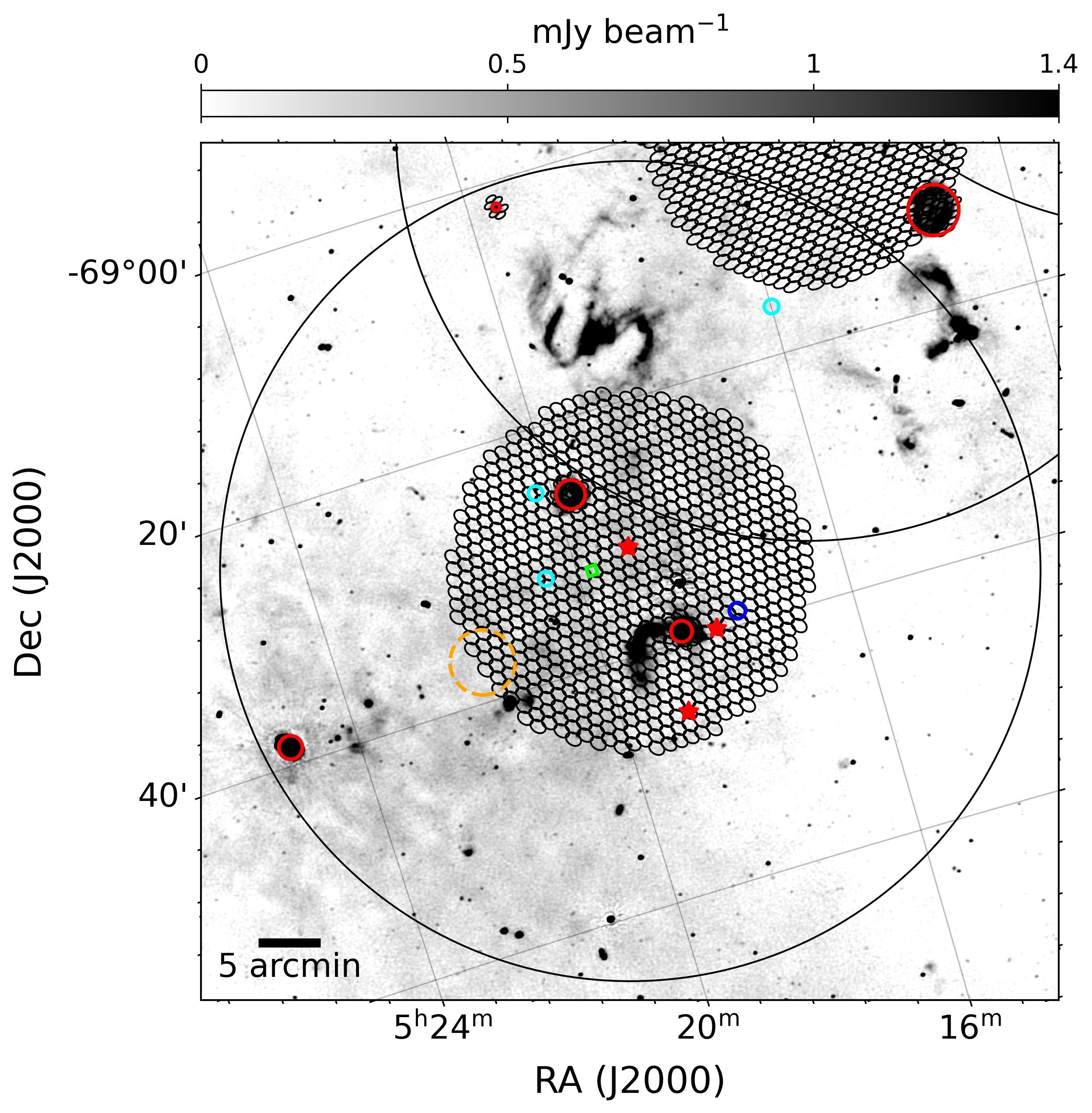} & \includegraphics[width=.46\linewidth]{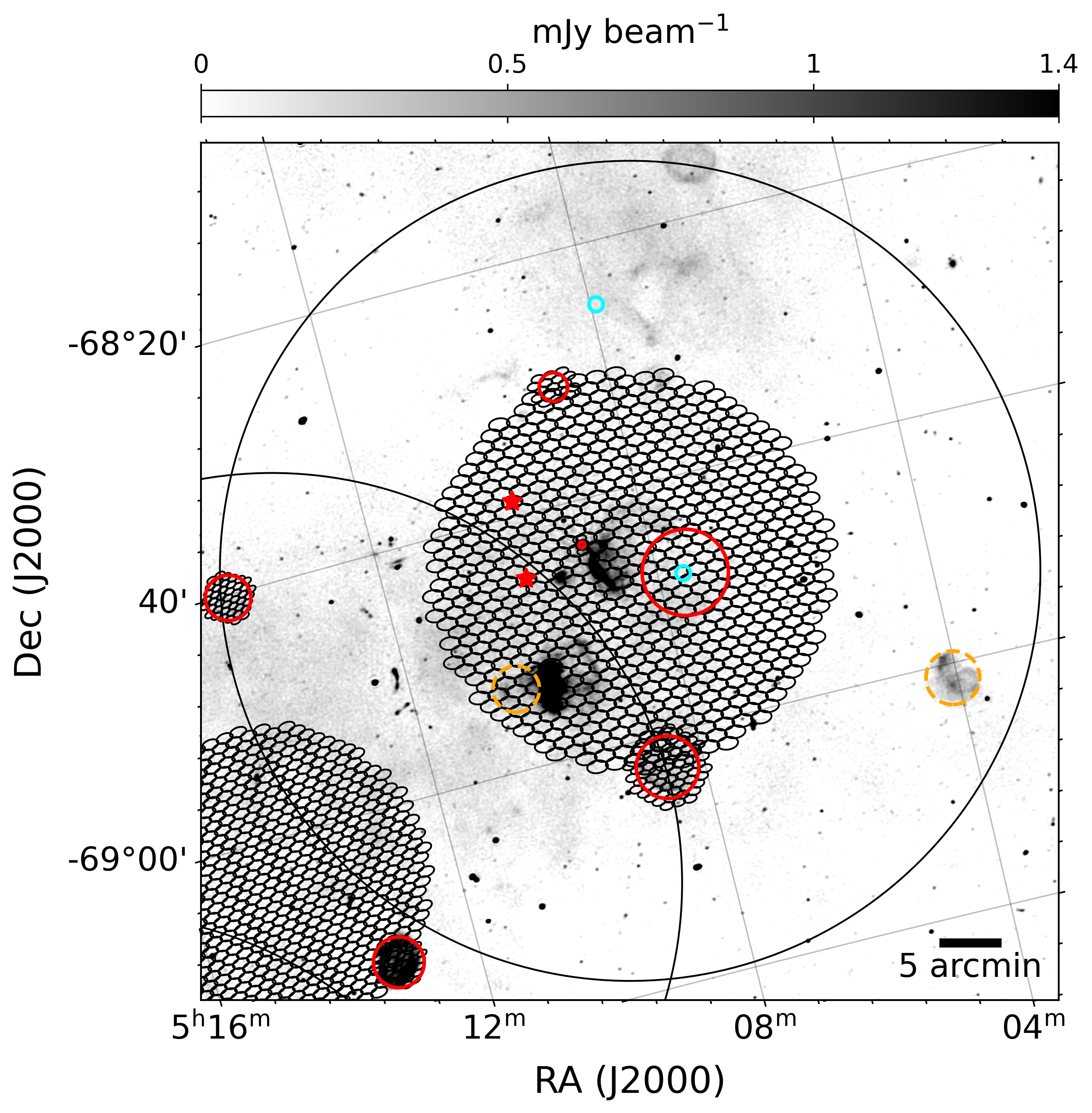}\\
\includegraphics[width=.46\linewidth]{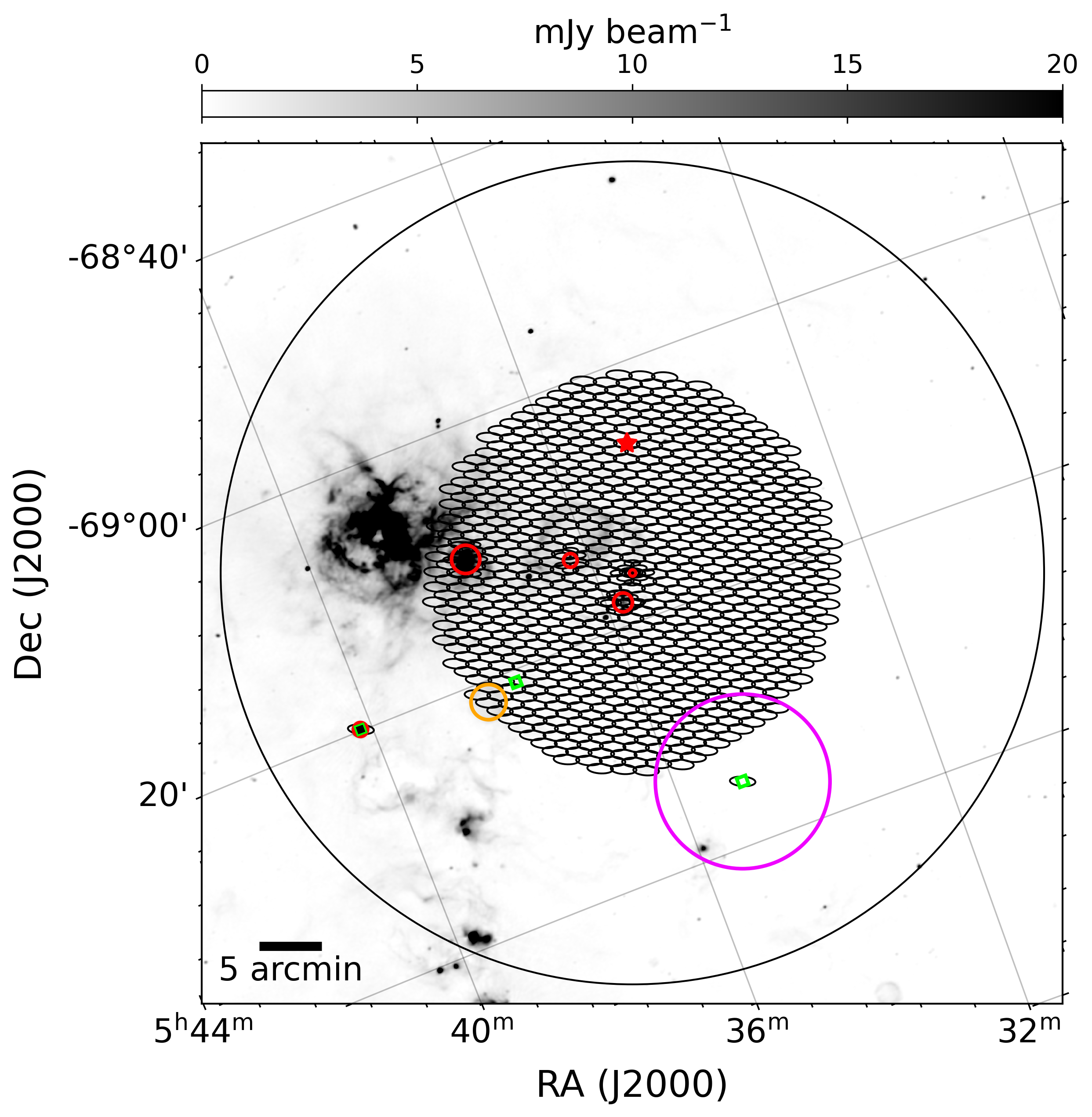} & \includegraphics[width=.46\linewidth]{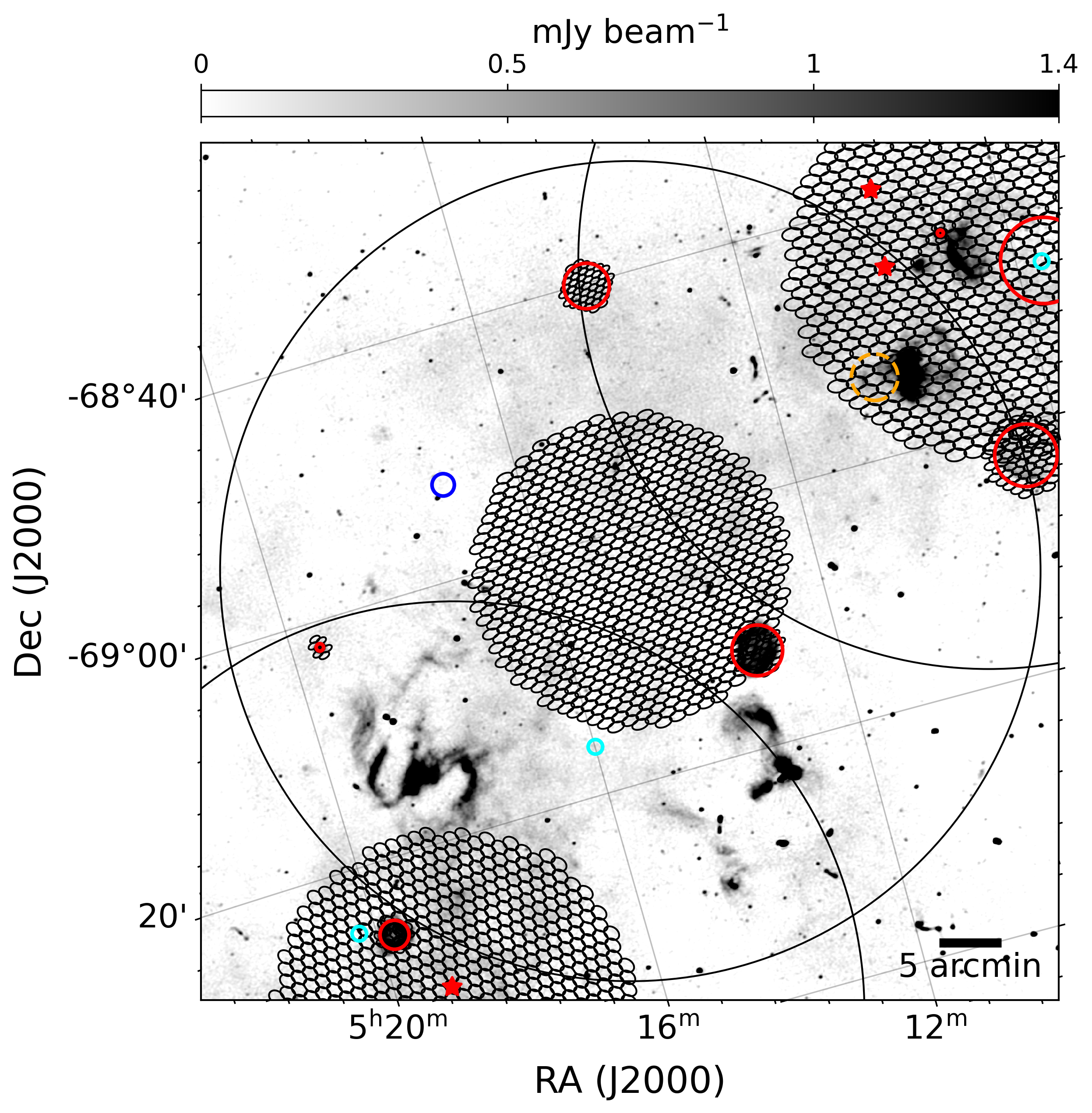}\\
\end{tabular}
\caption{Beam maps, made with the Python package \href{http://aplpy.github.io/}{\textsc{aplpy}} \protect\citep{Robitaille2012}, for Pointing\,1 (upper left), Pointing\,2 (upper right), Pointing\,3 (lower left), and Pointing\,4 (lower right), are shown overlaid onto an ASKAP 888\,MHz radio continuum survey image from \protect\cite{Pennock2021}. The outermost black circle is the size of the IB of MeerKAT at the HPBW at 1284\,MHz. The central tilings of CBs in black have a 50 per cent overlap, while the remaining tilings, depending on the source of interest, may have either a 50 or 70 per cent overlap (see \autoref{sec: Setup}). The red circles are SNRs, the green squares represent known pulsars, the blue circles are GCs, the cyan circles indicate HMXBs, and the orange circles are candidate SNRs (see \autoref{section: Source selection} and \autoref{tab:sources}). The dashed orange circles in Pointings 1 and 2 show candidate SNRs J0504$-$6901, J0510$-$6853, and J0521$-$6936 from \protect\cite{Bozzetto2023} and \protect\cite{Zangrandi2024}, which had not yet been discovered at the time of observation. In Pointing\,1, the SNR\,N\,132D, found at its bottom left, was not tiled due to its proximity to the HPBW of the IB. The pink circle in Pointing\,3 represents the error on the position of the known pulsar PSR\,J0535$-$6935 \citep{Crawford2001}. We did not detect this pulsar, possibly because it could be located outside of the CB tilings. The IB of Pointing\,3 was centred on the SN\,1987A remnant. The red star markers represent the new pulsar discoveries excluding the incoherent beam pulsar.} 
\label{fig:beam maps}
\end{figure*}
\end{center}

The first discovery of a pulsar in the LMC, PSR\,J0529$-$6652, and in fact the first extragalactic radio pulsar discovery, was made by \cite{McCulloch1983}. The discovery was part of a survey of the Magellanic Clouds using Murriyang, undertaken between 1980 and 1987, that was eventually improved and extended by \cite{McConnell1991}, yielding two more pulsars in the LMC. The observations lasted 5000\,s each, with the most common observation parameters being a centre frequency of 610\,MHz and a total bandwidth of 60\,MHz spread across 24 channels. The implementation of the Murriyang multibeam receiver \citep{Staveley-Smith1996} paved the way for higher sensitivity surveys such as the Parkes Multibeam Pulsar Survey (PMPS), which focused on the Galactic plane \citep{Manchester2001}, conducted between 2000 and 2001. Around the same time, using the Murriyang multibeam receiver, \cite{Manchester2006} searched for radio pulsars in the Magellanic Clouds. The most recent, and still ongoing, LMC survey using Murriyang, initiated in the year 2009, is known as the High Resolution LMC Survey \citep{Ridley2013, Hisano2022}.

\begin{figure}
\includegraphics[width=\columnwidth]{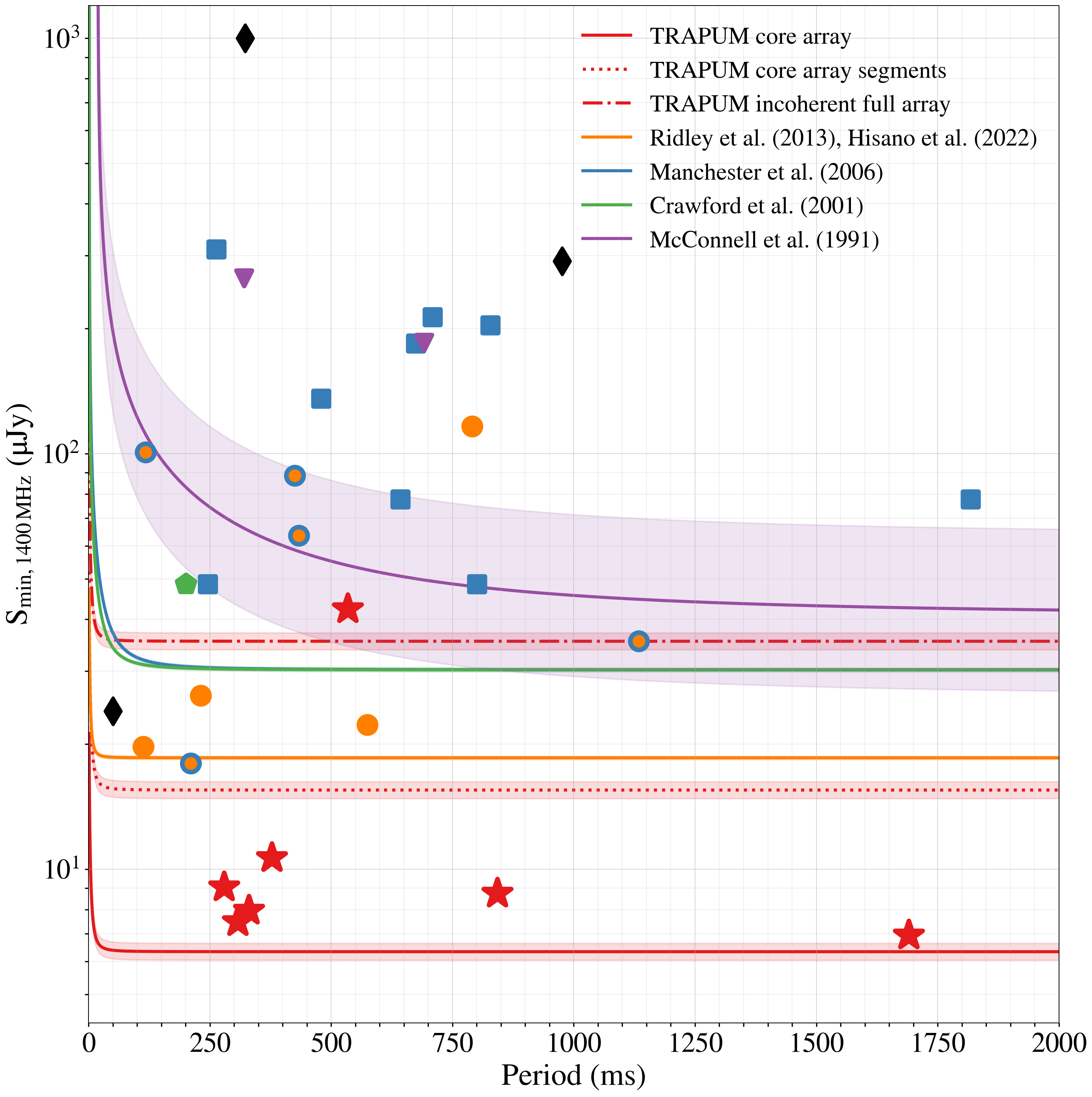}
\caption{The minimum flux densities at 1400\,MHz of the PMPS, the High Resolution LMC Survey, and this work are plotted against pulse period. The shaded regions around the sensitivity curves indicate the ranges of minimum flux density fluctuations, reflecting the pulsar power-law spectral index uncertainty as per \protect\cite{Jankowski2018}, which is $-1.6 \pm 0.54$. The minimum flux densities using Murriyang are slightly better than the literature, but this is based on our best knowledge of the surveys (see \autoref{sec: Survey sensitivity}). We note that the horizontal section of each of the minimum density curves excludes the red noise contribution, which is known to impact slow pulsars at longer integrations \citep[e.g.][]{Lazarus2015, vanHeerden2017}. Each of the markers, denoting the 24 radio pulsars discovered in the LMC, has a colour corresponding to the publication where the discovery was documented. The black diamond markers represent the flux densities for PSRs J0523$-$7125 \protect\citep{Wang2022}, J0529$-$6652 \protect\citep{Crawford2001}, and J0540$-$6919 \protect\citep{Johnston2004}. Notably, these pulsars have not been discovered in the main LMC pulsar surveys outlined in this paper. There are five orange circle markers with a blue outline representing pulsars which were reported in \protect\cite{Ridley2013} but discovered in the reprocessed archival PMPS data. One of these 5 blue-outlined orange circles, PSR\,J0535$-$66, has a 1400\,MHz flux density below the minimum flux densities of the Murriyang surveys, which is due to it being discovered using an integration time of 14400\,s \protect\citep{Ridley2013}. The newly discovered pulsars are represented by the red star markers, with the incoherent beam pulsar having the highest flux density (see \autoref{section: Discussion}).}
\label{fig:sensitivity}
\end{figure}

\cite{Manchester2006} observed the LMC using a total of 136 pointings, amounting to 1768 beams. Each pointing was observed for a duration of 8400\,s, with a sampling time of 1000\,$\upmu s$, a centre frequency of 1374\,MHz, and a bandwidth of 288\,MHz spread across 96 channels. The High Resolution LMC Survey kept the same centre frequency and improved on its predecessor by increasing the length of observation, the bandwidth, and the number of channels to 8600\,s, 340\,MHz, and 1024 channels respectively. The sampling time was also reduced to 64\,$\upmu s$ following the installation of the Berkeley Parkes Swinburne Recorder (BPSR, \citealt{Keith2010}).

A total of 13 pulsars were discovered in the LMC from the PMPS by \cite{Manchester2006}, with five found from a reprocessing of the data later on by \cite{Ridley2013}. With 54 out of 136 pointings completed, the High Resolution LMC survey has yielded four new radio pulsars so far \citep{Ridley2013, Hisano2022}. 

The `Crab twin' pulsar, PSR\,J0540$-$6919 \citep{Seward1984}, and PSR\,J0537$-$6910 \citep{Marshall1998} were discovered in X-ray observations. PSR\,J0540$-$6919 was subsequently observed across various wavelengths, including optical \citep{MiddleditchPennypacker1985}, radio \citep{Manchester1993}, and in gamma rays \citep{Marshall2016}, while PSR\,J0537$-$6910 remains exclusively detectable at X-rays. Interestingly, \cite{Crawford2001} made the serendipitous discovery of PSR\,J0535$-$6935 while trying to observe radio signals from PSR\,J0537$-$6910. The detection of PSR\,J0535$-$6935 was in observation data of lengths 21600\,s and 17200\,s. The only currently known radio pulsar in the LMC for which radio pulsations have not been discovered by Murriyang is PSR\,J0523$-$7125, found in the Australian Square Kilometre Array Pathfinder (ASKAP) Variables and Slow Transients survey \citep{Wang2022}, with both MeerKAT and Murriyang used afterwards to confirm the discovery. \autoref{tab:LMC pulsars} shows the properties of the known LMC pulsar population.

\begin{table}
\centering
\caption{Table adapted from \protect\cite{Ridley2013} showing the list of the 25 known LMC rotation-powered pulsars along with their spin periods, P, their fluxes at 1400\,MHz, ${\it S}_{\text{1400\,MHz}}$, their dispersion measures, DM, and their respective discovery papers \protect\citep{Manchester2005}. The ${\it S}_{\text{1400\,MHz}}$ values, which are not necessarily discovery flux densities, were obtained from \protect\cite{Ridley2013}, \protect\cite{Xie2019}, \protect\cite{Wang2022}, \protect\cite{Hisano2022}, and \protect\cite{Johnston2022}. The flux densities for eight pulsars are quoted as lower limits as these are not well localised, as reported in \protect\cite{Ridley2013}. PSR\,J0537$-$6910 is known to solely emit X-ray pulsations.}
\label{tab:LMC pulsars}
\begin{tabular}{l@{\hskip 0.13in}c@{\hskip 0.08in}c@{\hskip 0.08in}c@{\hskip 0.09in}r}
\hline
\textbf{Pulsar} & \textbf{P} & \boldmath{${\it S}_{\text{1400\,MHz}}$} & \textbf{DM} & \textbf{Discovery paper} \\
\text{JName} & \text{(ms)} & \text{{($\upmu Jy$)}} & \text{($pc \, cm^{-3}$)} & \text{} \\
\hline
J0449$-$7031 & 479.16 & 56 & 65.83 & \text{\cite{Manchester2006}} \\
J0451$-$67 & 245.45 & 50 & 45 & \text{\cite{Manchester2006}}\\
J0455$-$6951 & 320.42 & 83 & 94.70 & \text{\cite{McConnell1991}}\\
J0456$-$69 & 117.07 & $>$ 150  & 103 & \text{\cite{Ridley2013}}\\
J0456$-$7031 & 800.13 & 13 & 100.3 & \text{\cite{Manchester2006}}\\
J0457$-$69 & 231.39 & $>$ 50 & 91 & \text{\cite{Ridley2013}}\\
J0458$-$67 & 1133.9 & $>$ 70 & 97 & \text{\cite{Ridley2013}}\\
J0502$-$6617 & 691.25 & 250 & 68.9 & \text{\cite{McConnell1991}}\\
J0519$-$6932 & 263.21 & 130 & 118.86 & \text{\cite{Manchester2006}}\\
J0521$-$68  & 433.42 & $>$ 120 & 136 & \text{\cite{Ridley2013}}\\
J0522$-$6847 & 674.53 & 83 & 126.45 & \text{\cite{Manchester2006}}\\
J0523$-$7125 & 322.50 & 1000 & 157.5 & \text{\cite{Wang2022}}\\
J0529$-$6652 & 975.74 & 213 & 103.31 & \text{\cite{McCulloch1983}}\\
J0532$-$6639 & 642.74 & 42 & 69.2 & \text{\cite{Manchester2006}}\\
J0532$-$69 & 1149.2 & $>$ 50 & 124 & \text{\cite{Ridley2013}}\\
J0534$-$6703 & 1817.6 & 116 & 95.3 & \text{\cite{Manchester2006}}\\
J0535$-$66 & 210.52 & $>$ 30 & 75 & \text{\cite{Ridley2013}}\\
J0535$-$6935 & 200.51 & 50 & 93.7 & \text{\cite{Crawford2001}}\\
J0537$-$69 & 112.61 & $>$ 40 & 273 & \text{\cite{Ridley2013}}\\
J0537$-$6910 & 16.122 & - & - & \text{\cite{Marshall1998}}\\
J0540$-$6919 & 50.570 & 100 & 147.2 & \text{\cite{Seward1984}}\\
J0542$-$68 & 425.19  & $>$ 140 & 114 & \text{\cite{Ridley2013}}\\
J0543$-$6851 & 708.95 & 87 & 134.9 & \text{\cite{Manchester2006}}\\
J0555$-$7056 & 827.84 & 58 & 72.9 & \text{\cite{Manchester2006}}\\
J0556$-$67 & 790.55  & 120 & 71  & \text{\cite{Hisano2022}}\\
\hline
\end{tabular}
\end{table}

\section{Source selection}\label{section: Source selection}

To optimise our search strategies, we compiled an extensive list of known sources in the LMC that are likely to harbour pulsars. These included SNRs and candidate SNRs \citep{Bozzetto2017, Maitra2019, Yew2021, Maitra2021, Kavanagh2022, Sasaki2022, Bozzetto2023, Zangrandi2024}, HMXBs \citep{Antoniou2016, Maitra2019, Maitra2021, Haberl2022}, globular clusters (GCs, The SIMBAD astronomical database\footnote{\href{http://simbad.u-strasbg.fr/simbad/}{http://simbad.u-strasbg.fr/simbad/}}, \citealt{Wenger2000}), and pulsar wind nebulae (PWNe, \citealt{Manchester1993b, Gotthelf2000, Wang2001, Gaensler2003, Williams2005, Bamba2006, Haberl2012}). Below we describe the motivation behind each of these types of astrophysical targets. 

     \subsection{Supernova remnants and candidate supernova remnants}

    Given that NSs originate from the aftermath of massive stars going supernova \citep{BaadeZwicky1934}, it is intuitive to seek out pulsars within SNRs. Pulsars associated with SNRs are typically young, with examples such as the Crab pulsar and PSR\,J0540$-$6919 which are known to emit giant pulses \citep{Johnston2003, Geyer2021}. There are more than 60 pulsars that are associated with a SNR \citep{Manchester2005}. These remnants are distinguished by specific characteristics, including a non-thermal power-law radio spectral index steeper than $-0.4$, diffuse X-ray emissions, and a $[Sii]/H\alpha$ ratio equal to or greater than 0.4. Often, if a source has just one of these features, it is categorised as a candidate SNR \citep{Bozzetto2023}. The LMC hosts more than 80 SNRs and 40 candidate SNRs \citep{Zangrandi2024} and to improve our chances of discovering pulsars, we chose to target some of these. 

    One of these LMC SNRs is arguably the most famous recent supernova that we know of, SN\,1987A, a type II supernova located on the fringe of the most energetic star-forming region in the Local Group, 30\,Doradus \citep{Marchi2011, Fahrion2024}. Since the first detection of neutrinos emanating from SN\,1987A \citep{Bionta1987, Burrows1987, Hirata1987}, and despite decades of multi-wavelength searches \citep[e.g.][]{Manchester1996, Marshall1998, Zhang2018, Alp2018}, the existence of a pulsar in the SNR remains to be proven. \cite{Zanardo2018} found some indications for the existence of a compact object when analysing polarisation data, while other studies have detected signs of a PWN in the SN\,1987A remnant in both radio \citep{Cigan2019}, using the Atacama Large Millimeter/submillimeter Array (ALMA), and X-ray frequencies, using the \textit{Chandra}, \textit{NuSTAR}, and \textit{XMM-Newton} telescopes \citep{Greco2022}. The \textit{James Webb Space Telescope} (\textit{JWST}, \citealt{Gardner2006}) provided further insight, peeking through the dust and gas clouds of the SNR at infrared wavelengths, with \cite{Fransson2024} finding emission lines from highly ionised argon and sulphur which they attributed to being produced by a NS. Regarding pulsed radio emission from the SN\,1987A remnant, the current best limit is from \cite{Zhang2018}, who derived an upper limit of 31\,$\upmu Jy$ at 1400\,MHz by using Murriyang and making use of a signal-to-noise ratio (S/N) threshold of 8. We find that the sky temperature was not taken into account for this upper limit value. If we do include the sky temperature in the direction of the LMC in our calculation, which is 4.0\,K at the frequency of 1400\,MHz according to the Global Sky Model \citep{Zheng2017} as in the Python package \textsc{PyGDSM}\footnote{\href{https://github.com/telegraphic/pygdsm}{https://github.com/telegraphic/pygdsm} accessed on 2023 October 3} \citep{Price2016}, we expect an upper limit of 37.2\,$\upmu Jy$. Motivated by both this upper limit and our survey sensitivity (see \autoref{sec: Survey sensitivity}), we chose to conduct a fresh search for a radio pulsar in the SN\,1987A remnant using MeerKAT.

\begin{table}
\centering
\caption{List of sources targeted in the four pointings, excluding known radio pulsars.}
\label{tab:sources}
\begin{tabular}{lcc}
\hline
\textbf{Object} & \textbf{Type} & \textbf{Pointing}      \\ \hline

\begin{tabular}[c]{@{}l@{}} N\,120A \\ LHG\,27 \\ N\,132D \\ OGLE-CL\,LMC\,318 \\ XMMU\,J052016.0$-$692505 \\ LXP8.04 \end{tabular} & \begin{tabular}[c]{@{}c@{}} SNR \\ SNR \\ SNR \\ GC \\ HMXB \\ HMXB \end{tabular} & Pointing\,1 \\
\hline

\begin{tabular}[c]{@{}l@{}} HP\,791 \\ MCSNR\,J0508$-$6830 \\ N\,103B \\ MCSNR\,J0507$-$6847 \\ RX\,J050736$-$6847.8 \\ LXP169 \end{tabular} & \begin{tabular}[c]{@{}c@{}} SNR \\ SNR \\ SNR \\ SNR \\ HMXB \\ HMXB \end{tabular} & Pointing\,2 \\
\hline

\begin{tabular}[c]{@{}l@{}} SN\,1987A \\ Honeycomb \\ MCSNR\,J0536$-$6913 \\ 30\,Dor\,B \\ N\,158A \\ MCSNR\,J0538$-$6921 \\ PSR\,J0537$-$6910 \end{tabular} & \begin{tabular}[c]{@{}c@{}} SNR \\ SNR \\ SNR \\ SNR \\ SNR \\ Candidate SNR \\ PSR \end{tabular} & Pointing\,3 \\
\hline

\begin{tabular}[c]{@{}l@{}} N\,112 \\ LHG\,26 \\ HP\,700 \\ BRHT\,33b \\ RXJ0516.0$-$6916 \end{tabular} & \begin{tabular}[c]{@{}c@{}} SNR \\ SNR \\ SNR \\ GC \\ HMXB \end{tabular} & Pointing\,4 \\
\hline

\end{tabular}
\end{table}

    \subsection{High-mass X-ray binaries}
    
    A NS can be recycled through the accretion of matter via Roche-lobe overflow in the case of a low-mass X-ray binary (LMXB), or through stellar winds from a companion star in the case of a HMXB, resulting in X-ray emission during what is known as the X-ray binary phase. In a LMXB comprised of a NS and a low-mass companion, material is drawn from the companion star and the NS can be spun up into the millisecond spin period regime \citep{Alpar1982, Radhakrishnan1982, Tauris2006, Papitto2013}. Accretion processes are expected to alter magnetic field structures, or suppress the emission mechanism responsible for generating radio emission in NSs. Consequently, radio pulsations are generally not expected at this evolutionary stage \citep{Fornasini2023, Eijnden2024}. Most pulsars found in HMXBs are exclusively X-ray pulsars, and over 80 of them have been discovered in the Milky Way \citep{Vitaliy2023}. A few radio pulsars have been discovered in systems believed to have evolved from HMXBs \citep[e.g.][]{Johnston1992, Kaspi1994}. With around 50 HMXBs discovered in the LMC, we chose to target some of these sources in view of finding unexpected pulsed radio emission \citep{Antoniou2016, Maitra2019, Haberl2022}.
    
   \subsection{Globular clusters}

    GCs exhibit a significantly larger pulsar population per unit of stellar mass than the Milky Way disc, estimated to be approximately one hundred times greater and predominantly comprised of MSPs \citep{Clark1975, Katz1975, Abbate2021}. This arises due to the conducive conditions of high stellar densities in GCs, where there is a higher probability for isolated NSs forming binaries and being recycled to MSPs than elsewhere in the Milky Way. With over 150 identified Milky Way GCs, over 300 pulsars have been discovered within 44 of these clusters through various surveys\footnote{\href{https://www3.mpifr-bonn.mpg.de/staff/pfreire/GCpsr.html}{https://www3.mpifr-bonn.mpg.de/staff/pfreire/GCpsr.html} accessed on 2024 July 11} \citep[e.g.,][]{Ransom2005, Pan2021}, with a significant proportion discovered by MeerKAT \citep[e.g.,][]{Ridolfi2021, Abbate2022, Ridolfi2022, Chen2023}. Notably, the LMC is home to more than a dozen GCs having ages surpassing 11\,Gyr \citep{Piatti2019}, increasing the likelihood of hosting recycled pulsars. Therefore, we targeted some LMC GCs to enhance our chances of discovering binary systems, including MSP binaries.
    
    \subsection{Pulsar wind nebulae}
    
    PWNe are created through the interaction between outflowing winds emitted by a pulsar and the surrounding interstellar material and they can reside within SNRs. The presence of a pulsar within a SNR can be inferred by observing the interactions of its pulsar wind with the surrounding environment \citep[e.g.][]{Scargle1969, Bietenholz2004, Gaensler2002a}. Two LMC pulsars are located within PWNe, namely PSRs\,J0540$-$6919 \citep{Manchester1993b, Gotthelf2000}, and J0537$-$6910 \citep{Wang2001}. Additionally, there are three PWNe in the LMC without associated pulsars. These are situated in SNRs\,LHG\,1 \citep{Gaensler2003}, DEM\,N206 \citep{Williams2005}, and L241 \citep{Bamba2006}. We aim to observe these PWNe to detect potential associated pulsars over the course of our survey of the LMC.

\section{Observations}\label{section: Observations}
 
     \subsection{Setup}\label{sec: Setup}

    We used the relative positions and the size of the MeerKAT primary beam, the incoherent beam (hereafter IB) which is about 1.1 degrees in diameter at the half-power beam width (HPBW) at the L-band centre frequency \citep{Asad2021}, to group the selected sources together. The pointing centres were fine-tuned to encompass the maximum number of candidate sources within the IB (as described in \autoref{section: Source selection}). Thereafter, we used the \textsc{MOSAIC} software \citep{Chen2021} to simulate coherent beam (hereafter CB) maps, consisting of up to 768 beams (see later in this section), for our planned observations, and tiled the sources depending on their apparent sizes in the sky. The central region of the IB, where the sensitivity is highest, was tiled with the remaining CBs that were not targeting any specific source. Typically, to be more sensitive to pulsars, we used a 70 per cent overlap between neighbouring beams for SNRs and 50 per cent for remaining sources. An overlap value of 50 per cent implies that the neighbouring beams intersect at the point where their sensitivity is reduced to half of their maximum. \autoref{tab:sources} shows the LMC sources observed in the first four pointings of the survey. We also targeted known radio pulsars that fall within the four pointings, namely PSRs J0519$-$6932, J0535$-$6935, J0537$-$69, and J0540$-$6919. The detections of these pulsars give us an indication of the performance of the survey pipeline, as well as information about radio frequency interference (RFI) present during the observations. \autoref{fig:beam maps} shows the beam tilings for the four observed pointings. 
    
    The TRAPUM backend was used to record, store, and analyse the data taken for each observation. It comprises two user-supplied equipment (USE) backends provided by the Max-Planck-Institut-für-Radioastronomie (MPIfR) and the Max-Planck-Gesellschaft, namely the Filterbanking Beamformer User Supplied Equipment (FBFUSE, \citealt{Barr2018}) and the Accelerated Pulsar Search User Supplied Equipment \citep[APSUSE, see][and references therein]{Padmanabh2023}. FBFUSE is an advanced beamformer that can produce digitally steerable CBs in real-time, and it is set up to record data in multiples of 4 antennas at a time to be efficient. The data are captured in a filterbank format by APSUSE, a high-end computing cluster that shares its storage with FBFUSE, where they are stored. Each USE has nodes that are capable of both GPU-based and CPU-based processing. We used the core of MeerKAT, comprising 44 antennas clustered over a 1\,km radius, as it provides a favourable trade-off between sensitivity and beam coverage \citep{Chen2021}. Depending on availability, either 40 or 44 antennas were used for our observations. The observations were carried out using the L-band receiver, with frequencies 856--1712\,MHz, a bandwidth of 856\,MHz, and a central frequency of 1284\,MHz. The four pointings were observed on 2022 September 08 (pointings 1 and 2) and on 2023 January 20 (pointings 3 and 4). The data were recorded with 2048 frequency channels and a sampling time of 153\,$\upmu s$, enabling FBFUSE to form up to 768 `tied-array' CBs. A summary of the observation parameters is given in \autoref{tab:observation parameters}.
    
\begin{table*}
\centering
\caption{List of parameters for the four MeerKAT pointings. The coherent beams' major and minor axes are given in arcseconds, along with the position angles, which are measured from East to North on the sky \citep{Chen2021, {Bezuidenhout2023}}.}
\label{tab:observation parameters}
\begin{tabular}{ccccccc}
\hline
\textbf{Observation Date} & \multicolumn{1}{l}{\textbf{Pointing centre}} & \textbf{CB dishes (number of CBs)} & \textbf{IB dishes} & \textbf{CB size} & \textbf{Observation duration}\\ \hline

\begin{tabular}[c]{@{}c@{}}Pointing\,1\\ 2022 Sept 08\end{tabular} & \begin{tabular}[c]{@{}c@{}}05$^{\rm h}$19$^{\rm m}$13\fs00\\ $-$69\textdegree{}33\arcmin20\farcs00\end{tabular} & 40 (759) & 59 & 39.66\arcsec, 29.20\arcsec, $-$50.37$^{\circ}$ & 7172.50\,s\\ \hline

\begin{tabular}[c]{@{}c@{}}Pointing\,2\\ 2022 Sept 08 \end{tabular} & \begin{tabular}[c]{@{}c@{}}05$^{\rm h}$08$^{\rm m}$23\fs40\\ $-$68\textdegree{}46\arcmin35\farcs00\end{tabular} & 40 (762)& 59 & 50.33\arcsec, 30.40\arcsec, $-$4.40$^{\circ}$ & 7176.89\,s\\ \hline

\begin{tabular}[c]{@{}c@{}}Pointing\,3\\ 2023 Jan 20 \end{tabular} & \begin{tabular}[c]{@{}c@{}}05$^{\rm h}$35$^{\rm m}$28\fs02\\ $-$69\textdegree{}16\arcmin11\farcs10\end{tabular} & 44 (764) & 62 & 61.70\arcsec, 22.99\arcsec, $-$24.09$^{\circ}$ & 7160.58\,s\\ \hline

\begin{tabular}[c]{@{}c@{}}Pointing\,4\\ 2023 Jan 20 \end{tabular} & \begin{tabular}[c]{@{}c@{}}05$^{\rm h}$14$^{\rm m}$46\fs19\\ $-$69\textdegree{}03\arcmin19\farcs20\end{tabular} & 44 (766) & 62 & 41.27\arcsec, 23.16\arcsec, 7.82$^{\circ}$ & 7178.77\,s\\ \hline

\end{tabular}
\end{table*}

    \subsection{Survey sensitivity}\label{sec: Survey sensitivity}

    Compared to Murriyang, MeerKAT has approximately four times the antenna gain at $\sim$2.8\,K\,Jy$^{-1}$ \citep{Bailes2020} for the full array. To assess the differences in sensitivity among various LMC surveys, we employ the modified radiometer equation (\autoref{eqn: radiometer}). This delineates a telescope's sensitivity in detecting radio signals, providing the minimum detectable flux density, ${\it S}_{min}$, for a pulsar with a spin period, \textit{P}, and effective pulse width, \textit{W}. It is formulated as \citep{Dewey1985, Lorimer2004}:

    \begin{equation}
    {\it S}_{min} = \frac{{\betait} \: S/N \: ({\it T}_{sys}+{\it T}_{sky})}  {\:{\it G} \sqrt{{\it n}_{pol}\:{\it t}_{obs}\:\Delta \nu}} \sqrt{\frac{\it W}{\it {P-W}}}.
    \label{eqn: radiometer}
    \end{equation}

    For our survey, the S/N of the detected pulse is taken to be 9.5, which is our minimum spectral S/N cut (see \autoref{sec: Search}), the system temperature contribution, ${\it T}_{sys}$, is $18\,K$ \citep{Bailes2020}, the sky temperature contribution, ${\it T}_{sky}$, is $4.6\,K$ \citep{Price2016, Zheng2017}, the sensitivity degradation factor due to the signal digitisation, $\betait$, is $1.0$ for 8-bit digitisation \citep{Kouwenhoven2001}, ${\it G}$ is the antenna gain taken to be 1.84\,$K\,Jy^{-1}$ for MeerKAT's core, ${\it n}_{pol}=2$ is the number of orthogonal polarisations to be summed, ${\it t}_{obs}$ is the observing time, and $\Delta\nu$ is the total bandwidth.
    
    \autoref{fig:sensitivity} shows a comparison between the sensitivity curves of previous LMC pulsar surveys and this survey overlaid with the discovery flux densities of known LMC radio pulsars at 1400\,MHz (${\it S}_{\text{1400\,MHz}}$). The S/N thresholds were found from their respective papers, or assumed using our best knowledge of the surveys. We used a folded S/N threshold of 10 for \cite{McConnell1991}, a spectral S/N threshold of 7 for both \cite{Crawford2001} and \cite{Manchester2006}, and a folded S/N threshold of 8 for \cite{Ridley2013} and \cite{Hisano2022}. Therefore, the sensitivity of each survey does not scale directly with the telescope parameters. We made use of a Fast Fourier Transform (FFT) efficiency factor of 0.7 to convert between spectral S/N and folded S/N \citep{Morello2020}. We used a sensitivity degradation factor due to digitisation of 1.06 for the High Resolution LMC Survey (\citealt{Ridley2013} and \citealt{Hisano2022}), which makes use of BPSR's 2-bit digitisation, and 1.25 for surveys using 1-bit digitisation (\citealt{McConnell1991}, \citealt{Crawford2001}, and \citealt{Manchester2006}). All of the Murriyang surveys had a gain assumed to be 0.735\,$K\,Jy^{-1}$ \citep{Manchester2001}, and a sky temperature of 4.1\,K, except for \cite{McConnell1991}, which had a sky temperature of 17.8\,K due to a different centre frequency of 610\,MHz, compared to 1374\,MHz for the others \citep{Price2016, Zheng2017}. We assumed a radio power-law spectral index of $-1.6 \pm 0.54$ \citep{Jankowski2018} for converting the minimum flux densities of the surveys to ${\it S}_{\text{1400\,MHz}}$. The shaded regions in \autoref{fig:sensitivity} capture the uncertainty range corresponding to the radio power-law spectral index. The system temperature was assumed to be 44\,K based on \cite{McCulloch1983} for \cite{McConnell1991}, and 21\,K according to \cite{Manchester2001} for both \cite{Crawford2001} and \cite{Manchester2006}. For the High Resolution LMC Survey, the system temperature was taken to be 23\,K \citep{Keith2010}. Where the effective pulse widths and pulsar flux densities for the different surveys were not available in the literature, we assumed an intrinsic pulse width of 2.5 per cent of the pulsar period and considered additional contributions to the observed pulse width to obtain the flux densities. These included the DM smearing, ${\it \tau}_{DM}$, the survey sampling time, ${\it \tau}_{samp}$, and the scattering time-scale, ${\it \tau}_{scatt}$. The latter was based on the model by \cite{Bhat2004}. We used the median DM of the known LMC pulsars of 98.7\,$pc \, cm^{-3}$ for these calculations and a radio power-law spectral index of $-1.6$ to convert the flux densities of the known pulsars to ${\it S}_{\text{1400\,MHz}}$. A 2-hour observation with the MeerKAT core achieves a deeper flux sensitivity limit than the PMPS and the High Resolution LMC Survey. For periods greater than 50\,ms, our survey achieves $\sim$3 times the sensitivity of the High Resolution LMC Survey, reaching down to a limiting flux density, i.e., the faintest signal that can be detected from a radio pulsar located at the centre of both the IB and a CB, of 6.3\,$\upmu Jy$ at 1400\,MHz. 

    \subsection{Search} \label{sec: Search}

    The TRAPUM pulsar search pipeline consisted of three processes: searching, filtering, and folding. The search process involved the removal of RFI using \texttt{filtool} from the PulsarX\footnote{\href{https://github.com/ypmen/PulsarX}{https://github.com/ypmen/PulsarX} accessed on 2022 September 12} suite \citep{men2023}. A de-dispersion plan was created using \texttt{DDplan.py} from the \textsc{PRESTO} package \citep{Ransom2011}. We chose a DM range of 35--500\,$pc\,cm^{-3}$, given that for the known pulsars in the LMC, the lowest DM is 45\,$pc\,cm^{-3}$ and the highest is 273\,$pc\,cm^{-3}$ (see \autoref{tab:LMC pulsars}). This de-dispersion plan, with step sizes of 0.1 up to a DM of 334\,$pc\,cm^{-3}$, and 0.2 at higher DM values, was fed into the search pipeline where \textsc{PEASOUP}\footnote{\href{https://github.com/ewanbarr/peasoup}{https://github.com/ewanbarr/peasoup} accessed on 2022 September 12} \citep{Barr2020}, a GPU-based acceleration search software suite, performed the de-dispersion and the periodicity search \citep{Morello2019}. We configured \textsc{PEASOUP} to search for candidates with periods up to a maximum of 10\,s. To increase our sensitivity to pulsars located in binaries with shorter orbital periods where the latter can induce Doppler shifts on the signals causing pulse smearing, the 2-hour observation for each pointing was divided into six segments of 20-min each and a linear acceleration search was performed on each segment. For pointings 1 and 2, the acceleration search range was chosen to be $-$50 to 50\,$ms^{-2}$, with an acceleration tolerance of ten per cent for the segment searches, while for pointings 3 and 4, to cut down on processing time, we did not perform any acceleration search. The acceleration tolerance parameters ensure the pulse smearing due to acceleration step size stays within ten per cent of total smearing effects \citep{Levin2012}. The spectral S/N threshold of the FFT was set to be 8 with a maximum of 8 harmonics summed.

\begin{figure}
\includegraphics[width=\columnwidth]{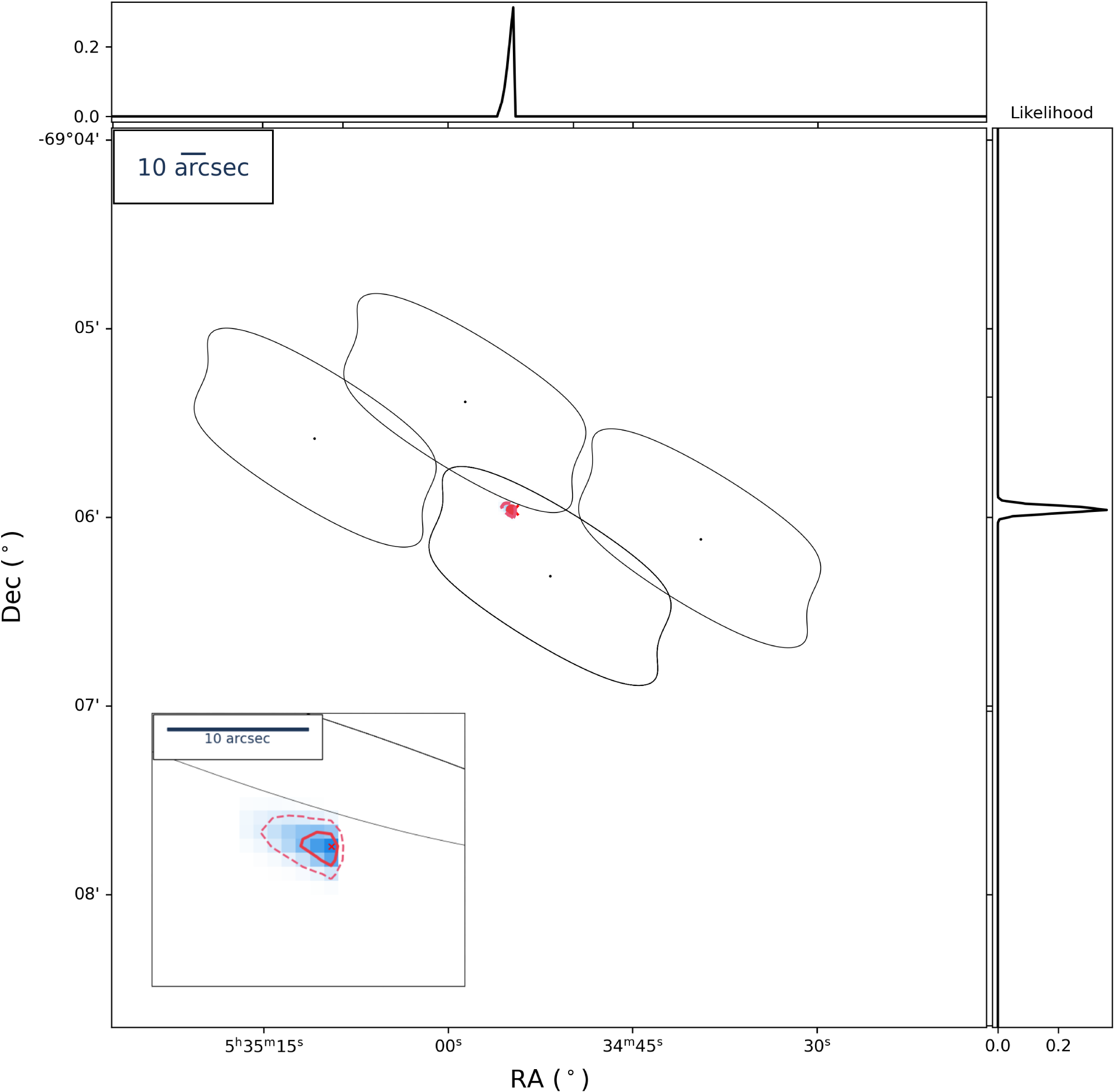}
\caption{The localisation of PSR\,J0534$-$6905 using \textsc{SeeKAT}, where the pulsar was detected in four beams, shown by the black outlines. The inset plot on the bottom left provides a zoomed-in view of the maximum-likelihood position, which is indicated by a red cross. The 1-$\sigma$ and 2-$\sigma$ errors on the position (see \autoref{tab:discoveries}) are demarcated by the red continuous line and the red dashed line, respectively.}
\label{fig:SeeKAT}
\end{figure}

After completing the search, we subjected the candidates to a filtering process using \textsc{candidate\_filter}\footnote{\href{https://github.com/prajwalvp/candidate\_filter}{https://github.com/prajwalvp/candidate\_filter} accessed on 2022 September 20}. This filter sifted known RFI signals and then clustered candidates considering spin periods, DMs, beam location, and acceleration values. Through this method, we effectively separated RFI signals, which tend to manifest across the entire field of view, from possible pulsar signals.

 \begin{table*}
\centering
\caption{Properties of the seven new pulsars, including the discovery flux density values (see \autoref{sec: Survey sensitivity}) and the pulse widths at 50 per cent of the peak intensity. The quoted uncertainties in parentheses are on the last digit. The error in each position is the 2-$\sigma$ output by \textsc{SeeKAT} (see \autoref{fig:SeeKAT}). For the pulsars denoted by a *, the error in their positions are the sizes of the CBs in which they were found (see \autoref{tab:observation parameters}). The RA and Dec of the incoherent beam pulsar are the same as the centre of Pointing\,1.}
\label{tab:discoveries}
\begin{tabular}{lllllcccc} 
\hline
\textbf{Pulsar} & \textbf{RA} & \textbf{Dec} & \textbf{DM} & \textbf{Period} & \textbf{Epoch} & \textbf{S/N} & \boldmath{${\it W}_{50}$} & \boldmath{${\it S}_{\text{1400\,MHz}}$} \\
\text{(J2000)} &  &  &  \text{($pc \, cm^{-3}$)} & \text{(ms)} & (MJD) &  & \text{(ms)} & \text{($\upmu Jy$)}\\
\hline
\addlinespace[0.1cm]

J0509$-$6838 & 05$^{\rm h}$09$^{\rm m}$58\fs17 $\substack{+3.0 \\ -1.8}$ & $-$68\textdegree{}38\arcmin46\farcs00 $\substack{+8.6 \\ -16.0}$ & 149(2) & 278.75363(26) & 59830.4 & 10.3 & $19.4(27)$ & 9.1\\ \addlinespace[0.2cm]

J0509$-$6845* & 05$^{\rm h}$09$^{\rm m}$55\fs26 & $-$68\textdegree{}45\arcmin09\farcs50 & 92.63(47) & 307.197131(77) & 59830.4 & 13.9 & $8.4(8)$ & 7.5\\ \addlinespace[0.2cm]

J0518$-$6939 & 05$^{\rm h}$18$^{\rm m}$15\fs50 $\substack{+1.0 \\ -0.7}$ & $-$69\textdegree{}39\arcmin45\farcs20 $\substack{+4.5 \\ -6.3}$ & 254.20(32) & 330.209968(56) & 59830.3 & 16.6 & $7.3(7)$ & 8.0\\ \addlinespace[0.2cm]

J0518$-$6946 & 05$^{\rm h}$18$^{\rm m}$58\fs39 $\substack{+1.1 \\ -1.6}$ & $-$69\textdegree{}46\arcmin16\farcs75 $\substack{+5.3 \\ -5.5}$ & 82(3) & 1690.4142(25) & 59830.3 & 13.9 & $40(3)$ & 6.9\\ \addlinespace[0.2cm]

J0519$-$6931* & 05$^{\rm h}$19$^{\rm m}$08\fs39   & $-$69\textdegree{}31\arcmin25\farcs60 & 82.70(26)  & 377.735862(52) & 59830.3 & 23.4 & $7.5(4)$ & 10.6\\ \addlinespace[0.2cm]

Incoherent beam & 05$^{\rm h}$19$^{\rm m}$13\fs00  & $-$69\textdegree{}33\arcmin20\farcs00 & 79(2) & 533.96689(51) & 59830.3 & 11.0 & $24(3)$ & 42.3\\ \addlinespace[0.2cm]

J0534$-$6905 & 05$^{\rm h}$34$^{\rm m}$54\fs69 $\substack{+0.0 \\ -0.9}$ & $-$69\textdegree{}05\arcmin57\farcs90 $\substack{+2.0 \\ -2.0}$ & 244(1) & 842.78252(58) & 59964.6 & 17.4 & $25(2)$ & 8.8
\\ \addlinespace[0.1cm]
\hline
\end{tabular}
\end{table*}

Following the filtering, the candidates were folded by making use of \texttt{psrfold\_fil}, a component of the \textsc{PulsarX} package. It used the DMs, spin periods, and acceleration values of the candidates derived from the search. Candidates with a spectral S/N above 9 were folded for pointings 1 and 2, while for pointings 3 and 4, a higher S/N threshold of 9.5 was used. This modification was made to optimise pipeline processing speed, reducing the number of candidates to fold, while maintaining our pulsar detection capabilities. We plan to keep the spectral S/N threshold as 9.5 for future pointings. Similarly, due to the high number of candidates, we applied a lowest period cut-off of 8 times the sampling time, i.e., 1.216\,$ms$. 

The output files containing pulsar candidates generated by the folding process were archive files that are compatible with the \textsc{PSRCHIVE} package \citep{Hotan2004, VanStraten2012}. These files were then classified using the Pulsar Image-based Classification System (\textsc{PICS}, \citealt{Zhu2014}), which employs deep neural networks for image pattern recognition in pulsar detection. PICS was trained using the Pulsar Arecibo L-band Feed Array (PALFA) survey data \citep{Cordes2006}, along with data from TRAPUM pulsar surveys \citep{Padmanabh2023}. PICS assigned a number ranging from 0 to 1 to the candidates, a proxy for probability, with a higher score implying that the candidate showed characteristics more akin to those of a pulsar. 

The resulting candidates having altogether a folded S/N above 7, a DM greater than 3, and a PICS score above 0.1 were then viewed using \textsc{CandyJar}\footnote{\href{https://github.com/vivekvenkris/CandyJar}{https://github.com/vivekvenkris/CandyJar} accessed on 2022 October 4}, enabling quick sorting and manual classification. Once we identified a candidate as an actual pulsar, we refolded the surrounding beams with the candidate parameters to verify additional detections for confirmation. The \textsc{SeeKAT} Python package\footnote{\href{https://github.com/BezuidenhoutMC/SeeKAT}{https://github.com/BezuidenhoutMC/SeeKAT} accessed on 2022 November 6} (hereafter \textsc{SeeKAT}, \citealt{Bezuidenhout2023}) improved localisation using these detections (see \autoref{fig:SeeKAT}), requiring a minimum of three detections to do so. It makes use of the S/N, the point spread function (PSF), and the location of the beams with detections to triangulate a position for a pulsar on the sky. We also refolded the detection beam of a pulsar using integer multiples of the pulsar period to potentially improve the discovery S/N ratio and accommodate for harmonics. A full description of the search pipeline can be found in \cite{Padmanabh2023} and \cite{Carli2024}.

\section{Results}\label{section: Results}

The first four pointings of the TRAPUM LMC Survey have yielded seven new pulsars, with four in Pointing\,1, two in Pointing\,2, and one in Pointing\,3. In Pointing\,1, one of the pulsars was only detected in the IB and therefore has poor localisation. We call this pulsar the incoherent beam pulsar, or IB pulsar for short, for the remainder of this work.

 The periods of the pulsars range from 278\,ms to 1690\,ms, categorising them as `slow' pulsars, or normal pulsars. They have DMs ranging from 79 to 254.2\,$pc \, cm^{-3}$, and if we consider the DM contribution from the Milky Way in the direction of the LMC, which is about 53\,$pc \, cm^{-3}$ using the NE2001 electron density model \citep{Cordes2004}, and about 58\,$pc \, cm^{-3}$ using the YMW16 electron density model \citep{Yao2017}, we can conclude that these pulsars are extragalactic and belong to the LMC. The DM range of the new pulsars is also consistent with the previously discovered LMC pulsars, which have DMs between 45\,$pc \, cm^{-3}$, which notably contradicts the estimations of the electron density models, and 273\,$pc \, cm^{-3}$ (see \autoref{tab:LMC pulsars}). We performed follow-up observations of the new pulsars discovered in pointings 1 and 2, and all of the pulsars were detected again except for the incoherent beam pulsar. The IBs of the follow-up observations were unfortunately not analysed to enable the confirmation of the incoherent beam pulsar, due to the requirement to rapidly delete data. It was also not detected in the IB of Pointing\,4, despite the latter overlapping with that of Pointing\,1 (see \autoref{fig:beam maps}) when processed using the search pipeline.
 All of the already known pulsars that were located within our pointings (see \autoref{sec: Setup}) were successfully detected, with the exception of PSR\,J0535$-$6935 (see \autoref{fig:beam maps}). We did not detect any pulsed radio emission from the observation of the SN\,1987A remnant and the X-ray pulsar PSR\,J0537$-$6910 in Pointing\,3 (see discussion in \autoref{section: Discussion}). 

 \begin{center}
\begin{figure*}
\begin{tabular}{c}
\includegraphics[width=2\columnwidth]{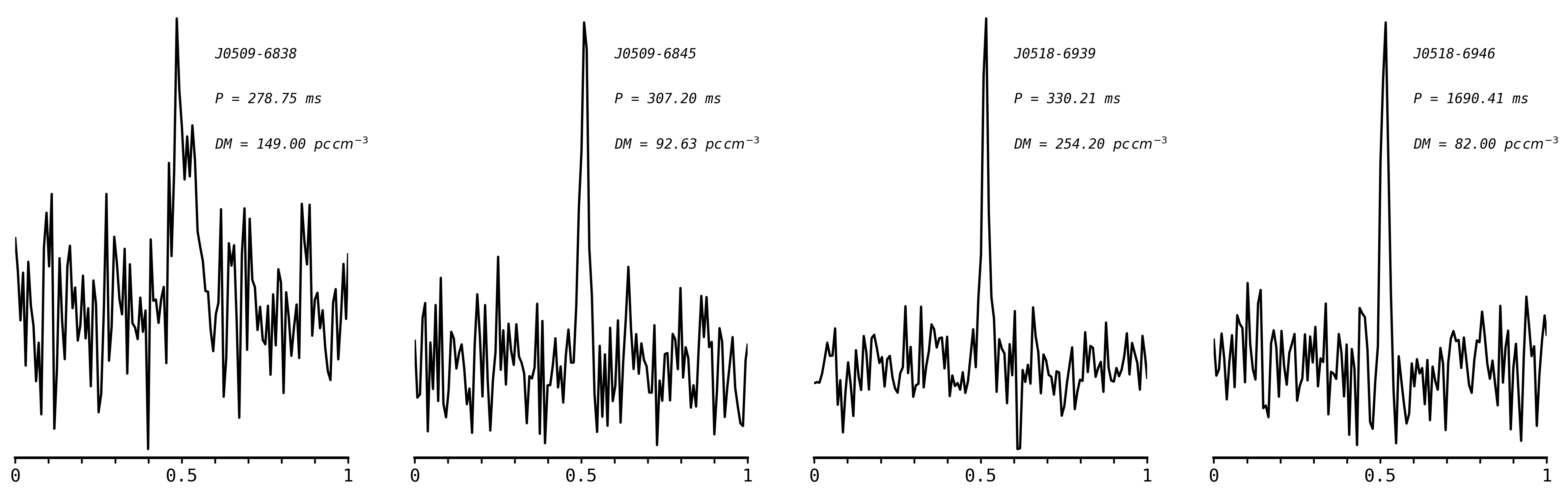} \\ \includegraphics[width=1.5\columnwidth]{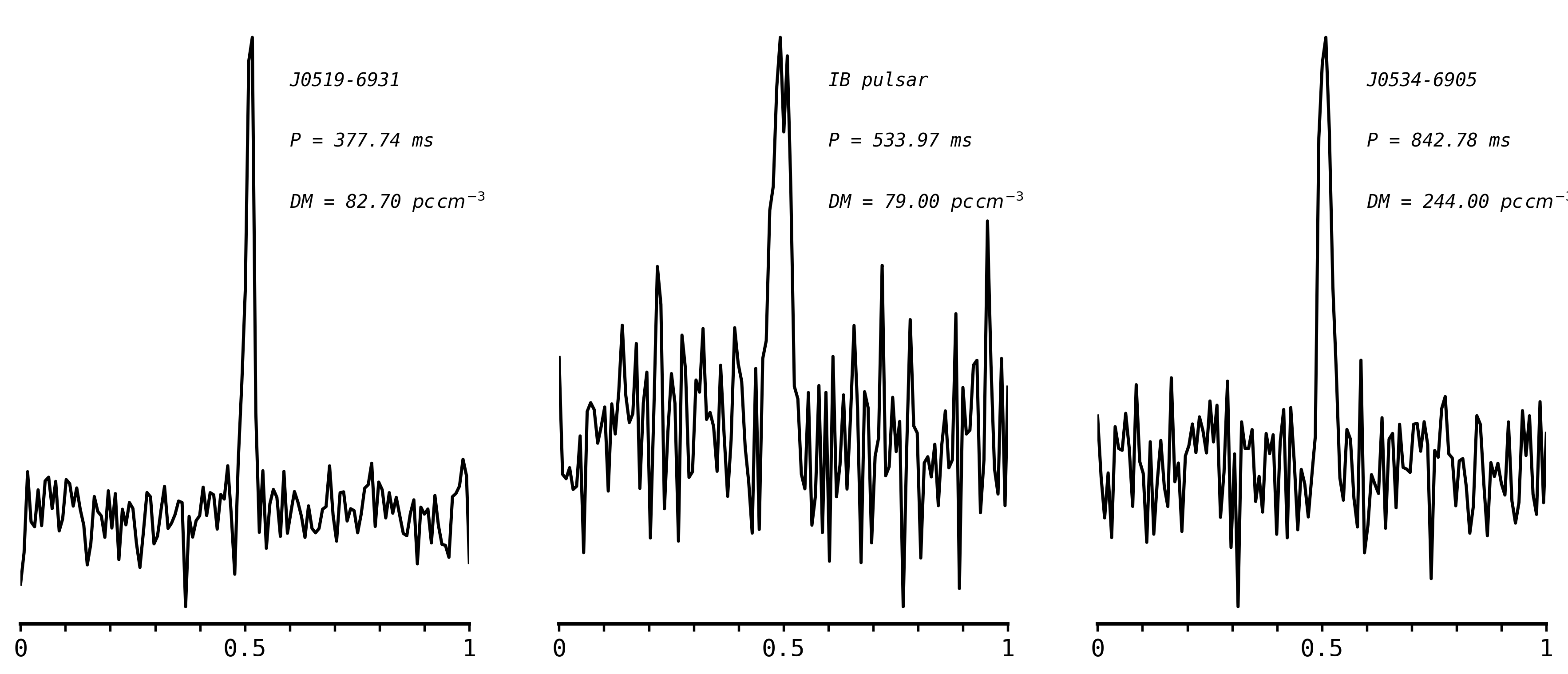}\\
\end{tabular}
\caption{The integrated pulse profiles, along with the periods and DMs, of the newly discovered pulsars within MeerKAT's L-band, with each plot representing one full rotation of the pulsar with 128 phase bins. The effective time resolution for each pulsar was less than its pulse width.}
\label{fig:pulse profiles}
\end{figure*}
\end{center}

 The integrated pulse profiles for the seven pulsars are shown in \autoref{fig:pulse profiles}, and they all have a single pulse profile. To determine the pulse width of each new discovery, we made a noise-less pulse template for each pulsar using the \texttt{paas} tool from \textsc{PSRCHIVE}, and we simulated 1000 noise profiles using the off-pulse noise statistics derived from their respective average pulse profiles. We fitted von Mises functions to each pulse profile for every simulated noise profile using the \texttt{fitvonMises} routine from \textsc{PSRSALSA}\footnote{\href{https://github.com/weltevrede/psrsalsa}{https://github.com/weltevrede/psrsalsa} accessed on 2023 October 25} \citep{Weltevrede2016}, which allowed us to measure the pulse widths at 50 per cent of the peak intensity, ${\it W}_{50}$, without being significantly influenced by noise. The ${\it W}_{50}$ for each pulsar was obtained by taking the average of the 1000 resultant pulse width measurements. \autoref{tab:discoveries} provides a summary of the properties of the new pulsars and their derived parameters.

\subsection{Discoveries} \label{section: Discoveries}
    \subsubsection{PSR\,J0509$-$6838}

    PSR\,J0509$-$6838 was discovered with a folded S/N of 10.3, obtained from PulsarX (see \autoref{sec: Search}), in the complete 2-hour observation data of Pointing\,2, making it the weakest detection out of the seven new discoveries. It was found within a beam of the central tiling of Pointing\,2 (see \autoref{fig:beam maps}), and further detected in the complete 2-hour data of two surrounding beams, enabling its localisation using \textsc{SeeKAT}. With a period of 278.75\,ms, it is the fastest rotating pulsar amongst the seven new discoveries. The ${\it W}_{50}$ is $(19.4 \pm 2.7)\,ms$, corresponding to a duty cycle of 7 per cent.

    \subsubsection{PSR\,J0509$-$6845}

    PSR\,J0509$-$6845 was discovered with a folded S/N of 13.9 in the complete 2-hour observation data of Pointing\,2. It was detected within a beam of the central tiling of Pointing\,2, with only one additional detection found in the 2-hour data of the surrounding beams. Consequently, we were unable to achieve a localisation more precise than the size of the coherent beam in which it was detected. The ${\it W}_{50}$ is $(8.4 \pm 0.8)\,ms$, resulting in a duty cycle of 2.7 per cent.

    \subsubsection{PSR\,J0518$-$6939}

    PSR\,J0518$-$6939 was discovered with a folded S/N of 16.6 in the complete 2-hour observation data of Pointing\,1. It was found within a beam of the central tiling of Pointing\,1 (see \autoref{fig:beam maps}), and further detected in the complete 2-hour data of two surrounding beams, enabling its localisation using \textsc{SeeKAT}. PSR\,J0518$-$6939 has a DM of 254.20\,$pc\,cm^{-3}$, which is the second-highest DM in the LMC, after PSR\,J0537$-$69 with a DM of 273\,$pc\,cm^{-3}$ (see \autoref{tab:LMC pulsars}). The ${\it W}_{50}$ is $(7.3 \pm 0.7)\,ms$, corresponding to a duty cycle of 2.2 per cent.

    \subsubsection{PSR\,J0518$-$6946}

    PSR\,J0518$-$6946 was discovered with a folded S/N of 13.9 in the complete 2-hour observation data of Pointing\,1. It was located within a beam of the central tiling. The pulsar was further detected in the complete 2-hour data of two surrounding beams, enabling its localisation using \textsc{SeeKAT}. PSR\,J0518$-$6946 has a period of 1.69\,s, making it the second-longest period radio pulsar in the Magellanic Clouds, following PSR\,J0534$-$6845 with a period of 1.82\,s (see \autoref{tab:LMC pulsars}). The ${\it W}_{50}$ is $(40 \pm 3)\,ms$, resulting in a duty cycle of 2.4 per cent.

    \subsubsection{PSR\,J0519$-$6931}

    PSR\,J0519$-$6931 was discovered with a folded S/N of 23.4 in the complete 2-hour observation data of Pointing\,1, making it the strongest detection amongst the seven new discoveries. It was found within a beam of the central tiling, with only one additional detection found in the 2-hour data of the surrounding beams. With a total of two detections, we were unable to obtain a better localisation than the size of the coherent beam in which it was detected. The ${\it W}_{50}$ is $(7.5 \pm 0.4)\,ms$, corresponding to a duty cycle of 2 per cent.

    \subsubsection{Incoherent beam pulsar}

    The incoherent beam pulsar was discovered with a folded S/N of 11.0 in the complete 2-hour observation data of Pointing\,1. It was located outside of the beam tilings of Pointing\,1, with a period of 533.97\,ms and a DM of 79\,$pc\,cm^{-3}$. The ${\it W}_{50}$ is $(24 \pm 3)\,ms$, resulting in a duty cycle of 4.5 per cent.

    \subsubsection{PSR\,J0534$-$6905} \label{section: Discoveries J0534-6905}
    
    PSR\,J0534$-$6905 was discovered with a folded S/N of 17.4 in the complete 2-hour observation data of Pointing\,3. It was found within a beam of the central tiling (see \autoref{fig:beam maps}), and further detected in the complete 2-hour data of three surrounding beams, enabling its localisation using \textsc{SeeKAT} (see \autoref{fig:SeeKAT}). The ${\it W}_{50}$ is $(25 \pm 2)\,ms$, corresponding to a duty cycle of 3 per cent.

\section{Discussion and conclusions}\label{section: Discussion}

The seven radio pulsars discovered increase the known LMC pulsar population to 32, an increase of $\sim$30 per cent. They do not seem to be associated with any of the sources listed in \autoref{tab:sources}. We cross-checked the positions with the ASKAP 888\,MHz radio continuum survey image of the LMC \citep{Pennock2021}, and we found that PSR\,J0518$-$6946 seems to be located within a faint shell of radio emission, possibly linked to the nearby SNR\,N\,120A, although this is not conclusive, with an on-sky angular separation of $\sim$432\,arcsec from the centre of the latter. Similarly, PSR\,J0518$-$6939 is also located in a region of radio emission, closer to SNR\,N\,120A. The on-sky angular separation in this case is $\sim$150\,arcsec. If PSR\,J0518$-$6939 is associated with SNR\,N\,120A, which has an estimated age of 7300--8000 years \citep{Rosado1993, Reyes2008}, it would have a maximum average transverse velocity of over 4000 km\,$s^{-1}$, one order of magnitude higher than the mean velocity of young pulsars, which we take to be $\sim$400 km\,$s^{-1}$ \citep{Lyne1994, Hobbs2005, Verbunt2017}, and several times higher than the fastest moving pulsars, around 1000 km\,$s^{-1}$ \citep{Chatterjee2005, Deller2019}, making the association unlikely. PSR\,J0534$-$6905 has a high DM of 244\,$pc\,cm^{-3}$, which can be explained by its location on the outskirts of the star-forming region, 30\,Doradus, where there is abundant gas and dust. Future LMC imaging surveys may reveal more about these pulsars and their associations.
 
The incoherent beam pulsar was detected with a ${\it S}_{\text{1400\,MHz}}$ of 42.3\,$\upmu Jy$ (see \autoref{tab:discoveries}), which implies it is bright enough to have been detectable in the High Resolution LMC Survey and the PMPS, as seen in \autoref{fig:sensitivity}. A detection within the previous surveys' data would help in framing the localisation and for this reason, we downloaded the Murriyang archival data available from the CSIRO Data Access Portal\footnote{\href{https://data.csiro.au/}{https://data.csiro.au/}} centred on the coordinates of the IB of Pointing\,1, and within a radius of 30\,arcmin. We undertook a search using \texttt{pdmp} from \textsc{PSRCHIVE} around the discovery DM, with a range of $\pm$10\,$pc \, cm^{-3}$, and discovery period, with a range of $\pm$10 per cent. Unfortunately, the searches did not yield any detection of the incoherent beam pulsar within the Murriyang archival data. This could be due to many factors, such as the distribution of the beams of the Murriyang surveys, and variations in the pulsar flux density.

The observation of the SN\,1987A remnant with MeerKAT was on day 13,115 after the explosion. The non-detection of radio pulsations from the SN\,1987A remnant could mean that the putative NS is still shrouded in substantial clouds of matter. The impact of the surrounding nebula varies significantly with observing frequency, the structure of surrounding dust and gas, and the viewing angle. With time, the ejected material is expected to clear out and could possibly give a better line of sight view of the suspected NS. The ejecta dust from the supernova has not gone through any reverse shock-related changes and it is expected that this could cause some dissipation when it does happen \citep{Jones2023}. We determine the upper limit for the detection of pulsed radio emission in the SN\,1987A remnant as 6.3\,$\upmu Jy$ at 1400\,MHz, since Pointing\,3 was centred on the SN\,1987A remnant (see \autoref{fig:beam maps}). We plan to re-observe the SN\,1987A remnant using the S-band receivers of MeerKAT, with frequencies 1.7–3.5\,GHz \citep{Barr2018}, to attempt to mitigate dispersion and scattering effects owing to the dust and gas clouds.

\cite{Crawford2024} detected three repeated signals from PSR\,J0537$-$6910 at a DM of 103.4\,$pc \, cm^{-3}$. We used \texttt{pdmp} to search the CB data from Pointing\,3 corresponding to the location of the X-ray pulsar around this DM value, with a range of $\pm 10\,pc \, cm^{-3}$, and the pulsar period with a range of $\pm10$ per cent. We unfortunately did not detect any pulsed radio emission from PSR\,J0537$-$6910, allowing us to derive the upper limit as 7.7\,$\upmu Jy$ at 1400\,MHz, after correcting for the sensitivity loss due to the offset between the location of the pulsar and the centre of Pointing\,3.

The seven new pulsars represent discoveries from the initial four pointings of the 28 currently planned for the TRAPUM LMC Survey conducted using MeerKAT. We save the low-resolution version of the data from all observed pointings for future Fast-Folding Algorithm searches. Follow-up timing observations are underway, and a detailed analysis will be presented in a future paper.

\section{Acknowledgements} \label{sec: Acknowledgements}
The MeerKAT telescope is operated by the South African Radio
Astronomy Observatory, which is a facility of the National Research Foundation, an agency of the Department of Science and Innovation. SARAO acknowledges the ongoing advice and calibration of GPS systems by the National Metrology Institute of South Africa (NMISA) and the time space reference systems department of the
Paris Observatory.

TRAPUM observations used the FBFUSE and APSUSE computing clusters for beamforming, data acquisition, storage, and analysis. These clusters were designed, funded, and installed by the Max-Planck-Institut für Radioastronomie and the Max-Planck-Gesellschaft. 

This research would not have been possible without the financial support for VP, which came from the National Organising Committee for the International Astronomical Union General Assembly 2024, through the Africa 2024 Scholarship, and the UCT-SKA Doctoral Scholarship. 

AP and MB are supported in part by the `Italian Ministry of Foreign Affairs and International Cooperation', grant number ZA23GR03, under the project `RADIOMAP - Science and technology pathways to MeerKAT+: the Italian and South African synergy'.

AR is supported by the Italian National Institute for Astrophysics (INAF) through an `IAF - Astrophysics Fellowship in Italy' fellowship (Codice Unico di Progetto: C59J21034720001; Project `MINERS'). AR also acknowledges continuing valuable support from the Max-Planck Society.

\section{Data availability}

The discovery PSRCHIVE archive files of the new pulsars can be found on \textsc{ZENODO} at \href{https://zenodo.org/doi/10.5281/zenodo.12723576}{DOI 10.5281/zenodo.12723576}. The TRAPUM collaboration will share the candidates list upon appropriate request. 

\bibliographystyle{mnras}
\bibliography{references} 

\begin{thebibliography}{}
\makeatletter
\relax
\def\mn@urlcharsother{\let\do\@makeother \do\$\do\&\do\#\do\^\do\_\do\%\do\~}
\def\mn@doi{\begingroup\mn@urlcharsother \@ifnextchar [ {\mn@doi@}
  {\mn@doi@[]}}
\def\mn@doi@[#1]#2{\def\@tempa{#1}\ifx\@tempa\@empty \href
  {http://dx.doi.org/#2} {doi:#2}\else \href {http://dx.doi.org/#2} {#1}\fi
  \endgroup}
\def\mn@eprint#1#2{\mn@eprint@#1:#2::\@nil}
\def\mn@eprint@arXiv#1{\href {http://arxiv.org/abs/#1} {{\tt arXiv:#1}}}
\def\mn@eprint@dblp#1{\href {http://dblp.uni-trier.de/rec/bibtex/#1.xml}
  {dblp:#1}}
\def\mn@eprint@#1:#2:#3:#4\@nil{\def\@tempa {#1}\def\@tempb {#2}\def\@tempc
  {#3}\ifx \@tempc \@empty \let \@tempc \@tempb \let \@tempb \@tempa \fi \ifx
  \@tempb \@empty \def\@tempb {arXiv}\fi \@ifundefined
  {mn@eprint@\@tempb}{\@tempb:\@tempc}{\expandafter \expandafter \csname
  mn@eprint@\@tempb\endcsname \expandafter{\@tempc}}}

\bibitem[\protect\citeauthoryear{Abbate}{Abbate}{2021}]{Abbate2021}
Abbate F.,  2021, Searching for pulsars in globular clusters with the MeerKAT
  Radio Telescope (\mn@eprint {arXiv} {2112.06528})

\bibitem[\protect\citeauthoryear{{Abbate} et~al.,}{{Abbate}
  et~al.}{2022}]{Abbate2022}
{Abbate} F.,  et~al., 2022, \mn@doi [\mnras] {10.1093/mnras/stac1041}, \href
  {https://ui.adsabs.harvard.edu/abs/2022MNRAS.513.2292A} {513, 2292}

\bibitem[\protect\citeauthoryear{{Agazie} et~al.,}{{Agazie}
  et~al.}{2023}]{Agazie2023}
{Agazie} G.,  et~al., 2023, \mn@doi [\apjl] {10.3847/2041-8213/acf4fd}, \href
  {https://ui.adsabs.harvard.edu/abs/2023ApJ...956L...3A} {956, L3}

\bibitem[\protect\citeauthoryear{{Alp} et~al.,}{{Alp} et~al.}{2018}]{Alp2018}
{Alp} D.,  et~al., 2018, \mn@doi [\apj] {10.3847/1538-4357/aad739}, \href
  {https://ui.adsabs.harvard.edu/abs/2018ApJ...864..174A} {864, 174}

\bibitem[\protect\citeauthoryear{{Alpar}, {Cheng}, {Ruderman}  \&
  {Shaham}}{{Alpar} et~al.}{1982}]{Alpar1982}
{Alpar} M.~A.,  {Cheng} A.~F.,  {Ruderman} M.~A.,   {Shaham} J.,  1982, \mn@doi
  [\nat] {10.1038/300728a0}, \href
  {https://ui.adsabs.harvard.edu/abs/1982Natur.300..728A} {300, 728}

\bibitem[\protect\citeauthoryear{{Antoniou} \& {Zezas}}{{Antoniou} \&
  {Zezas}}{2016}]{Antoniou2016}
{Antoniou} V.,  {Zezas} A.,  2016, \mn@doi [mnras] {10.1093/mnras/stw167},
  \href {https://ui.adsabs.harvard.edu/abs/2016MNRAS.459..528A} {459, 528}

\bibitem[\protect\citeauthoryear{{Asad} et~al.,}{{Asad}
  et~al.}{2021}]{Asad2021}
{Asad} K.~M.~B.,  et~al., 2021, \mn@doi [\mnras] {10.1093/mnras/stab104}, \href
  {https://ui.adsabs.harvard.edu/abs/2021MNRAS.502.2970A} {502, 2970}

\bibitem[\protect\citeauthoryear{{Baade} \& {Zwicky}}{{Baade} \&
  {Zwicky}}{1934}]{BaadeZwicky1934}
{Baade} W.,  {Zwicky} F.,  1934, \mn@doi [Proceedings of the National Academy
  of Science] {10.1073/pnas.20.5.254}, \href
  {https://ui.adsabs.harvard.edu/abs/1934PNAS...20..254B} {20, 254}

\bibitem[\protect\citeauthoryear{{Badenes}, {Maoz}  \& {Draine}}{{Badenes}
  et~al.}{2010}]{Badenes2010}
{Badenes} C.,  {Maoz} D.,   {Draine} B.~T.,  2010, \mn@doi [mnras]
  {10.1111/j.1365-2966.2010.17023.x}, \href
  {https://ui.adsabs.harvard.edu/abs/2010MNRAS.407.1301B} {407, 1301}

\bibitem[\protect\citeauthoryear{{Bailes} et~al.,}{{Bailes}
  et~al.}{2020}]{Bailes2020}
{Bailes} M.,  et~al., 2020, \mn@doi [pasa] {10.1017/pasa.2020.19}, \href
  {https://ui.adsabs.harvard.edu/abs/2020PASA...37...28B} {37, e028}

\bibitem[\protect\citeauthoryear{{Bamba}, {Ueno}, {Nakajima}, {Mori}  \&
  {Koyama}}{{Bamba} et~al.}{2006}]{Bamba2006}
{Bamba} A.,  {Ueno} M.,  {Nakajima} H.,  {Mori} K.,   {Koyama} K.,  2006,
  \mn@doi [\aap] {10.1051/0004-6361:20054096}, \href
  {https://ui.adsabs.harvard.edu/abs/2006A&A...450..585B} {450, 585}

\bibitem[\protect\citeauthoryear{{Barr}}{{Barr}}{2018}]{Barr2018}
{Barr} E.~D.,  2018, in {Weltevrede} P.,  {Perera} B.~B.~P.,  {Preston} L.~L.,
   {Sanidas} S.,  eds, Pulsar Astrophysics the Next Fifty Years. pp 175--178,
  \mn@doi{10.1017/S1743921317009036}

\bibitem[\protect\citeauthoryear{{Barr}}{{Barr}}{2020}]{Barr2020}
{Barr} E.,  2020, {Peasoup: C++/CUDA GPU pulsar searching library},
  Astrophysics Source Code Library, record ascl:2001.014 (\mn@eprint {ascl}
  {2001.014})

\bibitem[\protect\citeauthoryear{{Basu} et~al.,}{{Basu}
  et~al.}{2022}]{Basu2022}
{Basu} A.,  et~al., 2022, \mn@doi [\mnras] {10.1093/mnras/stab3336}, \href
  {https://ui.adsabs.harvard.edu/abs/2022MNRAS.510.4049B} {510, 4049}

\bibitem[\protect\citeauthoryear{{Bezuidenhout} et~al.,}{{Bezuidenhout}
  et~al.}{2023}]{Bezuidenhout2023}
{Bezuidenhout} M.~C.,  et~al., 2023, \mn@doi [RAS Techniques and Instruments]
  {10.1093/rasti/rzad007}, \href
  {https://ui.adsabs.harvard.edu/abs/2023RASTI...2..114B} {2, 114}

\bibitem[\protect\citeauthoryear{{Bhat}, {Cordes}, {Camilo}, {Nice}  \&
  {Lorimer}}{{Bhat} et~al.}{2004}]{Bhat2004}
{Bhat} N.~D.~R.,  {Cordes} J.~M.,  {Camilo} F.,  {Nice} D.~J.,   {Lorimer}
  D.~R.,  2004, \mn@doi [\apj] {10.1086/382680}, \href
  {https://ui.adsabs.harvard.edu/abs/2004ApJ...605..759B} {605, 759}

\bibitem[\protect\citeauthoryear{{Bietenholz}, {Hester}, {Frail}  \&
  {Bartel}}{{Bietenholz} et~al.}{2004}]{Bietenholz2004}
{Bietenholz} M.~F.,  {Hester} J.~J.,  {Frail} D.~A.,   {Bartel} N.,  2004,
  \mn@doi [apj] {10.1086/424653}, \href
  {https://ui.adsabs.harvard.edu/abs/2004ApJ...615..794B} {615, 794}

\bibitem[\protect\citeauthoryear{{Bionta} et~al.,}{{Bionta}
  et~al.}{1987}]{Bionta1987}
{Bionta} R.~M.,  et~al., 1987, \mn@doi [\prl] {10.1103/PhysRevLett.58.1494},
  \href {https://ui.adsabs.harvard.edu/abs/1987PhRvL..58.1494B} {58, 1494}

\bibitem[\protect\citeauthoryear{{Bozzetto} et~al.,}{{Bozzetto}
  et~al.}{2017}]{Bozzetto2017}
{Bozzetto} L.~M.,  et~al., 2017, \mn@doi [\apjs] {10.3847/1538-4365/aa653c},
  \href {https://ui.adsabs.harvard.edu/abs/2017ApJS..230....2B} {230, 2}

\bibitem[\protect\citeauthoryear{{Bozzetto} et~al.,}{{Bozzetto}
  et~al.}{2023}]{Bozzetto2023}
{Bozzetto} L.~M.,  et~al., 2023, \mn@doi [\mnras] {10.1093/mnras/stac2922},
  \href {https://ui.adsabs.harvard.edu/abs/2023MNRAS.518.2574B} {518, 2574}

\bibitem[\protect\citeauthoryear{{Burrows} \& {Lattimer}}{{Burrows} \&
  {Lattimer}}{1987}]{Burrows1987}
{Burrows} A.,  {Lattimer} J.~M.,  1987, \mn@doi [\apjl] {10.1086/184938}, \href
  {https://ui.adsabs.harvard.edu/abs/1987ApJ...318L..63B} {318, L63}

\bibitem[\protect\citeauthoryear{{Camilo} et~al.,}{{Camilo}
  et~al.}{2018}]{Camilo2018}
{Camilo} F.,  et~al., 2018, \mn@doi [\apj] {10.3847/1538-4357/aab35a}, \href
  {https://ui.adsabs.harvard.edu/abs/2018ApJ...856..180C} {856, 180}

\bibitem[\protect\citeauthoryear{{Carli} et~al.,}{{Carli}
  et~al.}{2024}]{Carli2024}
{Carli} E.,  et~al., 2024, \mn@doi [\mnras] {10.1093/mnras/stae1310}, \href
  {https://ui.adsabs.harvard.edu/abs/2024MNRAS.531.2835C} {531, 2835}

\bibitem[\protect\citeauthoryear{{Chatterjee} et~al.,}{{Chatterjee}
  et~al.}{2005}]{Chatterjee2005}
{Chatterjee} S.,  et~al., 2005, \mn@doi [\apjl] {10.1086/491701}, \href
  {https://ui.adsabs.harvard.edu/abs/2005ApJ...630L..61C} {630, L61}

\bibitem[\protect\citeauthoryear{{Chen}, {Barr}, {Karuppusamy}, {Kramer}  \&
  {Stappers}}{{Chen} et~al.}{2021}]{Chen2021}
{Chen} W.,  {Barr} E.,  {Karuppusamy} R.,  {Kramer} M.,   {Stappers} B.,  2021,
  \mn@doi [Journal of Astronomical Instrumentation]
  {10.1142/S2251171721500136}, \href
  {https://ui.adsabs.harvard.edu/abs/2021JAI....1050013C} {10, 2150013}

\bibitem[\protect\citeauthoryear{{Chen} et~al.,}{{Chen}
  et~al.}{2023}]{Chen2023}
{Chen} W.,  et~al., 2023, \mn@doi [\mnras] {10.1093/mnras/stad029}, \href
  {https://ui.adsabs.harvard.edu/abs/2023MNRAS.520.3847C} {520, 3847}

\bibitem[\protect\citeauthoryear{{Choudhury} et~al.,}{{Choudhury}
  et~al.}{2020}]{Choudhury2020}
{Choudhury} S.,  et~al., 2020, \mn@doi [\mnras] {10.1093/mnras/staa2140}, \href
  {https://ui.adsabs.harvard.edu/abs/2020MNRAS.497.3746C} {497, 3746}

\bibitem[\protect\citeauthoryear{{Choudhury} et~al.,}{{Choudhury}
  et~al.}{2021}]{Choudhury2021}
{Choudhury} S.,  et~al., 2021, \mn@doi [mnras] {10.1093/mnras/stab2446}, \href
  {https://ui.adsabs.harvard.edu/abs/2021MNRAS.507.4752C} {507, 4752}

\bibitem[\protect\citeauthoryear{{Cigan} et~al.,}{{Cigan}
  et~al.}{2019}]{Cigan2019}
{Cigan} P.,  et~al., 2019, \mn@doi [\apj] {10.3847/1538-4357/ab4b46}, \href
  {https://ui.adsabs.harvard.edu/abs/2019ApJ...886...51C} {886, 51}

\bibitem[\protect\citeauthoryear{{Clark}}{{Clark}}{1975}]{Clark1975}
{Clark} G.~W.,  1975, \mn@doi [\apjl] {10.1086/181869}, \href
  {https://ui.adsabs.harvard.edu/abs/1975ApJ...199L.143C} {199, L143}

\bibitem[\protect\citeauthoryear{{Cordes}}{{Cordes}}{2004}]{Cordes2004}
{Cordes} J.~M.,  2004, in {Clemens} D.,  {Shah} R.,   {Brainerd} T.,  eds,
  Astronomical Society of the Pacific Conference Series Vol. 317, Milky Way
  Surveys: The Structure and Evolution of our Galaxy. p.~211

\bibitem[\protect\citeauthoryear{{Cordes} et~al.,}{{Cordes}
  et~al.}{2006}]{Cordes2006}
{Cordes} J.~M.,  et~al., 2006, \mn@doi [apj] {10.1086/498335}, \href
  {https://ui.adsabs.harvard.edu/abs/2006ApJ...637..446C} {637, 446}

\bibitem[\protect\citeauthoryear{{Crawford}}{{Crawford}}{2024}]{Crawford2024}
{Crawford} F.,  2024, \mn@doi [\apj] {10.3847/1538-4357/ad5010}, \href
  {https://ui.adsabs.harvard.edu/abs/2024ApJ...968...99C} {968, 99}

\bibitem[\protect\citeauthoryear{{Crawford}, {Kaspi}, {Manchester}, {Lyne},
  {Camilo}  \& {D'Amico}}{{Crawford} et~al.}{2001}]{Crawford2001}
{Crawford} F.,  {Kaspi} V.~M.,  {Manchester} R.~N.,  {Lyne} A.~G.,  {Camilo}
  F.,   {D'Amico} N.,  2001, \mn@doi [apj] {10.1086/320635}, \href
  {https://ui.adsabs.harvard.edu/abs/2001ApJ...553..367C} {553, 367}

\bibitem[\protect\citeauthoryear{{De Marchi} et~al.,}{{De Marchi}
  et~al.}{2011}]{Marchi2011}
{De Marchi} G.,  et~al., 2011, \mn@doi [\apj] {10.1088/0004-637X/739/1/27},
  \href {https://ui.adsabs.harvard.edu/abs/2011ApJ...739...27D} {739, 27}

\bibitem[\protect\citeauthoryear{{Deller} et~al.,}{{Deller}
  et~al.}{2019}]{Deller2019}
{Deller} A.~T.,  et~al., 2019, \mn@doi [\apj] {10.3847/1538-4357/ab11c7}, \href
  {https://ui.adsabs.harvard.edu/abs/2019ApJ...875..100D} {875, 100}

\bibitem[\protect\citeauthoryear{{Dewey}, {Taylor}, {Weisberg}  \&
  {Stokes}}{{Dewey} et~al.}{1985}]{Dewey1985}
{Dewey} R.~J.,  {Taylor} J.~H.,  {Weisberg} J.~M.,   {Stokes} G.~H.,  1985,
  \mn@doi [\apjl] {10.1086/184502}, \href
  {https://ui.adsabs.harvard.edu/abs/1985ApJ...294L..25D} {294, L25}

\bibitem[\protect\citeauthoryear{{EPTA Collaboration} et~al.,}{{EPTA
  Collaboration} et~al.}{2023}]{Antoniadis2023}
{EPTA Collaboration} et~al., 2023, \mn@doi [\aap]
  {10.1051/0004-6361/202346844}, \href
  {https://ui.adsabs.harvard.edu/abs/2023A&A...678A..50E} {678, A50}

\bibitem[\protect\citeauthoryear{{Fahrion} \& {De Marchi}}{{Fahrion} \& {De
  Marchi}}{2024}]{Fahrion2024}
{Fahrion} K.,  {De Marchi} G.,  2024, \mn@doi [\aap]
  {10.1051/0004-6361/202348097}, \href
  {https://ui.adsabs.harvard.edu/abs/2024A&A...681A..20F} {681, A20}

\bibitem[\protect\citeauthoryear{{Fornasini}, {Antoniou}  \&
  {Dubus}}{{Fornasini} et~al.}{2023}]{Fornasini2023}
{Fornasini} F.,  {Antoniou} V.,   {Dubus} G.,  2023, in , Handbook of X-ray and
  Gamma-ray Astrophysics (eds. C. Bambi.
p.~143, \mn@doi{10.1007/978-981-16-4544-0_95-1}

\bibitem[\protect\citeauthoryear{Fransson et~al.,}{Fransson
  et~al.}{2024}]{Fransson2024}
Fransson C.,  et~al., 2024, \mn@doi [Science] {10.1126/science.adj5796}, 383,
  898

\bibitem[\protect\citeauthoryear{{Gaensler}, {Arons}, {Kaspi}, {Pivovaroff},
  {Kawai}  \& {Tamura}}{{Gaensler} et~al.}{2002}]{Gaensler2002a}
{Gaensler} B.~M.,  {Arons} J.,  {Kaspi} V.~M.,  {Pivovaroff} M.~J.,  {Kawai}
  N.,   {Tamura} K.,  2002, \mn@doi [apj] {10.1086/339354}, \href
  {https://ui.adsabs.harvard.edu/abs/2002ApJ...569..878G} {569, 878}

\bibitem[\protect\citeauthoryear{{Gaensler}, {Hendrick}, {Reynolds}  \&
  {Borkowski}}{{Gaensler} et~al.}{2003}]{Gaensler2003}
{Gaensler} B.~M.,  {Hendrick} S.~P.,  {Reynolds} S.~P.,   {Borkowski} K.~J.,
  2003, \mn@doi [\apjl] {10.1086/378687}, \href
  {https://ui.adsabs.harvard.edu/abs/2003ApJ...594L.111G} {594, L111}

\bibitem[\protect\citeauthoryear{{Gardner} et~al.,}{{Gardner}
  et~al.}{2006}]{Gardner2006}
{Gardner} J.~P.,  et~al., 2006, \mn@doi [\ssr] {10.1007/s11214-006-8315-7},
  \href {https://ui.adsabs.harvard.edu/abs/2006SSRv..123..485G} {123, 485}

\bibitem[\protect\citeauthoryear{{Geyer} et~al.,}{{Geyer}
  et~al.}{2021}]{Geyer2021}
{Geyer} M.,  et~al., 2021, \mn@doi [\mnras] {10.1093/mnras/stab1501}, \href
  {https://ui.adsabs.harvard.edu/abs/2021MNRAS.505.4468G} {505, 4468}

\bibitem[\protect\citeauthoryear{{Gotthelf} \& {Wang}}{{Gotthelf} \&
  {Wang}}{2000}]{Gotthelf2000}
{Gotthelf} E.~V.,  {Wang} Q.~D.,  2000, \mn@doi [\apjl] {10.1086/312568}, \href
  {https://ui.adsabs.harvard.edu/abs/2000ApJ...532L.117G} {532, L117}

\bibitem[\protect\citeauthoryear{{Graczyk} et~al.,}{{Graczyk}
  et~al.}{2020}]{Graczyk2020}
{Graczyk} D.,  et~al., 2020, \mn@doi [apj] {10.3847/1538-4357/abbb2b}, \href
  {https://ui.adsabs.harvard.edu/abs/2020ApJ...904...13G} {904, 13}

\bibitem[\protect\citeauthoryear{{Greco} et~al.,}{{Greco}
  et~al.}{2022}]{Greco2022}
{Greco} E.,  et~al., 2022, \mn@doi [\apj] {10.3847/1538-4357/ac679d}, \href
  {https://ui.adsabs.harvard.edu/abs/2022ApJ...931..132G} {931, 132}

\bibitem[\protect\citeauthoryear{{Grimm}, {Gilfanov}  \& {Sunyaev}}{{Grimm}
  et~al.}{2003}]{Grimm2003}
{Grimm} H.~J.,  {Gilfanov} M.,   {Sunyaev} R.,  2003, \mn@doi [\mnras]
  {10.1046/j.1365-8711.2003.06224.x}, \href
  {https://ui.adsabs.harvard.edu/abs/2003MNRAS.339..793G} {339, 793}

\bibitem[\protect\citeauthoryear{{Haberl} et~al.,}{{Haberl}
  et~al.}{2012}]{Haberl2012}
{Haberl} F.,  et~al., 2012, \mn@doi [\aap] {10.1051/0004-6361/201218971}, \href
  {https://ui.adsabs.harvard.edu/abs/2012A&A...543A.154H} {543, A154}

\bibitem[\protect\citeauthoryear{{Haberl}, {Maitra}, {Vasilopoulos}, {Maggi},
  {Udalski}, {Monageng}  \& {Buckley}}{{Haberl} et~al.}{2022}]{Haberl2022}
{Haberl} F.,  {Maitra} C.,  {Vasilopoulos} G.,  {Maggi} P.,  {Udalski} A.,
  {Monageng} I.~M.,   {Buckley} D.~A.~H.,  2022, \mn@doi [\aap]
  {10.1051/0004-6361/202243301}, \href
  {https://ui.adsabs.harvard.edu/abs/2022A&A...662A..22H} {662, A22}

\bibitem[\protect\citeauthoryear{{Haskell} \& {Melatos}}{{Haskell} \&
  {Melatos}}{2015}]{Haskell2015}
{Haskell} B.,  {Melatos} A.,  2015, \mn@doi [International Journal of Modern
  Physics D] {10.1142/S0218271815300086}, \href
  {https://ui.adsabs.harvard.edu/abs/2015IJMPD..2430008H} {24, 1530008}

\bibitem[\protect\citeauthoryear{{Hirata} et~al.,}{{Hirata}
  et~al.}{1987}]{Hirata1987}
{Hirata} K.,  et~al., 1987, \mn@doi [\prl] {10.1103/PhysRevLett.58.1490}, \href
  {https://ui.adsabs.harvard.edu/abs/1987PhRvL..58.1490H} {58, 1490}

\bibitem[\protect\citeauthoryear{{Hisano} et~al.,}{{Hisano}
  et~al.}{2022}]{Hisano2022}
{Hisano} S.,  et~al., 2022, \mn@doi [apj] {10.3847/1538-4357/ac5802}, \href
  {https://ui.adsabs.harvard.edu/abs/2022ApJ...928..161H} {928, 161}

\bibitem[\protect\citeauthoryear{{Hobbs}, {Lorimer}, {Lyne}  \&
  {Kramer}}{{Hobbs} et~al.}{2005}]{Hobbs2005}
{Hobbs} G.,  {Lorimer} D.~R.,  {Lyne} A.~G.,   {Kramer} M.,  2005, \mn@doi
  [\mnras] {10.1111/j.1365-2966.2005.09087.x}, \href
  {https://ui.adsabs.harvard.edu/abs/2005MNRAS.360..974H} {360, 974}

\bibitem[\protect\citeauthoryear{{Hotan}, {van Straten}  \&
  {Manchester}}{{Hotan} et~al.}{2004}]{Hotan2004}
{Hotan} A.~W.,  {van Straten} W.,   {Manchester} R.~N.,  2004, \mn@doi [\pasa]
  {10.1071/AS04022}, \href
  {https://ui.adsabs.harvard.edu/abs/2004PASA...21..302H} {21, 302}

\bibitem[\protect\citeauthoryear{{Jankowski}, {van Straten}, {Keane}, {Bailes},
  {Barr}, {Johnston}  \& {Kerr}}{{Jankowski} et~al.}{2018}]{Jankowski2018}
{Jankowski} F.,  {van Straten} W.,  {Keane} E.~F.,  {Bailes} M.,  {Barr} E.~D.,
   {Johnston} S.,   {Kerr} M.,  2018, \mn@doi [\mnras] {10.1093/mnras/stx2476},
  \href {https://ui.adsabs.harvard.edu/abs/2018MNRAS.473.4436J} {473, 4436}

\bibitem[\protect\citeauthoryear{{Johnston} \& {Romani}}{{Johnston} \&
  {Romani}}{2003}]{Johnston2003}
{Johnston} S.,  {Romani} R.~W.,  2003, \mn@doi [\apjl] {10.1086/376826}, \href
  {https://ui.adsabs.harvard.edu/abs/2003ApJ...590L..95J} {590, L95}

\bibitem[\protect\citeauthoryear{{Johnston}, {Manchester}, {Lyne}, {Bailes},
  {Kaspi}, {Qiao}  \& {D'Amico}}{{Johnston} et~al.}{1992}]{Johnston1992}
{Johnston} S.,  {Manchester} R.~N.,  {Lyne} A.~G.,  {Bailes} M.,  {Kaspi}
  V.~M.,  {Qiao} G.,   {D'Amico} N.,  1992, \mn@doi [\apjl] {10.1086/186300},
  \href {https://ui.adsabs.harvard.edu/abs/1992ApJ...387L..37J} {387, L37}

\bibitem[\protect\citeauthoryear{{Johnston}, {Romani}, {Marshall}  \&
  {Zhang}}{{Johnston} et~al.}{2004}]{Johnston2004}
{Johnston} S.,  {Romani} R.~W.,  {Marshall} F.~E.,   {Zhang} W.,  2004, \mn@doi
  [\mnras] {10.1111/j.1365-2966.2004.08286.x}, \href
  {https://ui.adsabs.harvard.edu/abs/2004MNRAS.355...31J} {355, 31}

\bibitem[\protect\citeauthoryear{{Johnston} et~al.,}{{Johnston}
  et~al.}{2022}]{Johnston2022}
{Johnston} S.,  et~al., 2022, \mn@doi [\mnras] {10.1093/mnras/stab3360}, \href
  {https://ui.adsabs.harvard.edu/abs/2022MNRAS.509.5209J} {509, 5209}

\bibitem[\protect\citeauthoryear{{Jonas} \& {MeerKAT Team}}{{Jonas} \& {MeerKAT
  Team}}{2016}]{Jonas2016}
{Jonas} J.,  {MeerKAT Team} 2016, in MeerKAT Science: On the Pathway to the
  SKA. p.~1, \mn@doi{10.22323/1.277.0001}

\bibitem[\protect\citeauthoryear{{Jones} et~al.,}{{Jones}
  et~al.}{2023}]{Jones2023}
{Jones} O.~C.,  et~al., 2023, \mn@doi [\apj] {10.3847/1538-4357/ad0036}, \href
  {https://ui.adsabs.harvard.edu/abs/2023ApJ...958...95J} {958, 95}

\bibitem[\protect\citeauthoryear{{Kaspi}, {Johnston}, {Bell}, {Manchester},
  {Bailes}, {Bessell}, {Lyne}  \& {D'Amico}}{{Kaspi} et~al.}{1994}]{Kaspi1994}
{Kaspi} V.~M.,  {Johnston} S.,  {Bell} J.~F.,  {Manchester} R.~N.,  {Bailes}
  M.,  {Bessell} M.,  {Lyne} A.~G.,   {D'Amico} N.,  1994, \mn@doi [\apjl]
  {10.1086/187231}, \href
  {https://ui.adsabs.harvard.edu/abs/1994ApJ...423L..43K} {423, L43}

\bibitem[\protect\citeauthoryear{{Katz}}{{Katz}}{1975}]{Katz1975}
{Katz} J.~I.,  1975, \mn@doi [\nat] {10.1038/253698a0}, \href
  {https://ui.adsabs.harvard.edu/abs/1975Natur.253..698K} {253, 698}

\bibitem[\protect\citeauthoryear{{Kavanagh}, {Sasaki}, {Filipovi{\'c}},
  {Points}, {Bozzetto}, {Haberl}, {Maggi}  \& {Maitra}}{{Kavanagh}
  et~al.}{2022}]{Kavanagh2022}
{Kavanagh} P.~J.,  {Sasaki} M.,  {Filipovi{\'c}} M.~D.,  {Points} S.~D.,
  {Bozzetto} L.~M.,  {Haberl} F.,  {Maggi} P.,   {Maitra} C.,  2022, \mn@doi
  [\mnras] {10.1093/mnras/stac813}, \href
  {https://ui.adsabs.harvard.edu/abs/2022MNRAS.515.4099K} {515, 4099}

\bibitem[\protect\citeauthoryear{{Keith} et~al.,}{{Keith}
  et~al.}{2010}]{Keith2010}
{Keith} M.~J.,  et~al., 2010, \mn@doi [mnras]
  {10.1111/j.1365-2966.2010.17325.x}, \href
  {https://ui.adsabs.harvard.edu/abs/2010MNRAS.409..619K} {409, 619}

\bibitem[\protect\citeauthoryear{{Kim}, {Izmailova}  \& {Aimuratov}}{{Kim}
  et~al.}{2023}]{Vitaliy2023}
{Kim} V.,  {Izmailova} I.,   {Aimuratov} Y.,  2023, \mn@doi [\apjs]
  {10.3847/1538-4365/ace68f}, \href
  {https://ui.adsabs.harvard.edu/abs/2023ApJS..268...21K} {268, 21}

\bibitem[\protect\citeauthoryear{{Kouwenhoven} \& {Vo{\^u}te}}{{Kouwenhoven} \&
  {Vo{\^u}te}}{2001}]{Kouwenhoven2001}
{Kouwenhoven} M.~L.~A.,  {Vo{\^u}te} J.~L.~L.,  2001, \mn@doi [\aap]
  {10.1051/0004-6361:20011226}, \href
  {https://ui.adsabs.harvard.edu/abs/2001A&A...378..700K} {378, 700}

\bibitem[\protect\citeauthoryear{{Lazarus} et~al.,}{{Lazarus}
  et~al.}{2015}]{Lazarus2015}
{Lazarus} P.,  et~al., 2015, \mn@doi [\apj] {10.1088/0004-637X/812/1/81}, \href
  {https://ui.adsabs.harvard.edu/abs/2015ApJ...812...81L} {812, 81}

\bibitem[\protect\citeauthoryear{{Levin}}{{Levin}}{2012}]{Levin2012}
{Levin} L.,  2012, PhD thesis, Swinburne University of Technology, Australia

\bibitem[\protect\citeauthoryear{{Liu}, {van Paradijs}  \& {van den
  Heuvel}}{{Liu} et~al.}{2005}]{Liu2005}
{Liu} Q.~Z.,  {van Paradijs} J.,   {van den Heuvel} E.~P.~J.,  2005, \mn@doi
  [\aap] {10.1051/0004-6361:20053718}, \href
  {https://ui.adsabs.harvard.edu/abs/2005A&A...442.1135L} {442, 1135}

\bibitem[\protect\citeauthoryear{{Lorimer} \& {Kramer}}{{Lorimer} \&
  {Kramer}}{2004}]{Lorimer2004}
{Lorimer} D.~R.,  {Kramer} M.,  2004, {Handbook of Pulsar Astronomy}.
Cambridge University Press

\bibitem[\protect\citeauthoryear{{Lyne}}{{Lyne}}{1999}]{Lyne1999}
{Lyne} A.,  1999, in {Arzoumanian} Z.,  {Van der Hooft} F.,   {van den Heuvel}
  E.~P.~J.,  eds, Pulsar Timing, General Relativity and the Internal Structure
  of Neutron Stars. p.~141

\bibitem[\protect\citeauthoryear{{Lyne} \& {Lorimer}}{{Lyne} \&
  {Lorimer}}{1994}]{Lyne1994}
{Lyne} A.~G.,  {Lorimer} D.~R.,  1994, \mn@doi [\nat] {10.1038/369127a0}, \href
  {https://ui.adsabs.harvard.edu/abs/1994Natur.369..127L} {369, 127}

\bibitem[\protect\citeauthoryear{{Maitra} et~al.,}{{Maitra}
  et~al.}{2019}]{Maitra2019}
{Maitra} C.,  et~al., 2019, \mn@doi [\mnras] {10.1093/mnras/stz2831}, \href
  {https://ui.adsabs.harvard.edu/abs/2019MNRAS.490.5494M} {490, 5494}

\bibitem[\protect\citeauthoryear{{Maitra}, {Haberl}, {Maggi}, {Kavanagh},
  {Vasilopoulos}, {Sasaki}, {Filipovi{\'c}}  \& {Udalski}}{{Maitra}
  et~al.}{2021}]{Maitra2021}
{Maitra} C.,  {Haberl} F.,  {Maggi} P.,  {Kavanagh} P.~J.,  {Vasilopoulos} G.,
  {Sasaki} M.,  {Filipovi{\'c}} M.~D.,   {Udalski} A.,  2021, \mn@doi [\mnras]
  {10.1093/mnras/stab716}, \href
  {https://ui.adsabs.harvard.edu/abs/2021MNRAS.504..326M} {504, 326}

\bibitem[\protect\citeauthoryear{{Manchester} \& {Peterson}}{{Manchester} \&
  {Peterson}}{1996}]{Manchester1996}
{Manchester} R.~N.,  {Peterson} B.~A.,  1996, \mn@doi [\apjl] {10.1086/309877},
  \href {https://ui.adsabs.harvard.edu/abs/1996ApJ...456L.107M} {456, L107}

\bibitem[\protect\citeauthoryear{{Manchester}, {Mar}, {Lyne}, {Kaspi}  \&
  {Johnston}}{{Manchester} et~al.}{1993a}]{Manchester1993}
{Manchester} R.~N.,  {Mar} D.~P.,  {Lyne} A.~G.,  {Kaspi} V.~M.,   {Johnston}
  S.,  1993a, \mn@doi [apjl] {10.1086/186714}, \href
  {https://ui.adsabs.harvard.edu/abs/1993ApJ...403L..29M} {403, L29}

\bibitem[\protect\citeauthoryear{{Manchester}, {Staveley-Smith}  \&
  {Kesteven}}{{Manchester} et~al.}{1993b}]{Manchester1993b}
{Manchester} R.~N.,  {Staveley-Smith} L.,   {Kesteven} M.~J.,  1993b, \mn@doi
  [\apj] {10.1086/172877}, \href
  {https://ui.adsabs.harvard.edu/abs/1993ApJ...411..756M} {411, 756}

\bibitem[\protect\citeauthoryear{{Manchester} et~al.,}{{Manchester}
  et~al.}{2001}]{Manchester2001}
{Manchester} R.~N.,  et~al., 2001, \mn@doi [mnras]
  {10.1046/j.1365-8711.2001.04751.x}, \href
  {https://ui.adsabs.harvard.edu/abs/2001MNRAS.328...17M} {328, 17}

\bibitem[\protect\citeauthoryear{{Manchester}, {Hobbs}, {Teoh}  \&
  {Hobbs}}{{Manchester} et~al.}{2005}]{Manchester2005}
{Manchester} R.~N.,  {Hobbs} G.~B.,  {Teoh} A.,   {Hobbs} M.,  2005, \mn@doi
  [aj] {10.1086/428488}, \href
  {https://ui.adsabs.harvard.edu/abs/2005AJ....129.1993M} {129, 1993}

\bibitem[\protect\citeauthoryear{{Manchester}, {Fan}, {Lyne}, {Kaspi}  \&
  {Crawford}}{{Manchester} et~al.}{2006}]{Manchester2006}
{Manchester} R.~N.,  {Fan} G.,  {Lyne} A.~G.,  {Kaspi} V.~M.,   {Crawford} F.,
  2006, \mn@doi [apj] {10.1086/505461}, \href
  {https://ui.adsabs.harvard.edu/abs/2006ApJ...649..235M} {649, 235}

\bibitem[\protect\citeauthoryear{{Marshall}, {Gotthelf}, {Zhang}, {Middleditch}
   \& {Wang}}{{Marshall} et~al.}{1998}]{Marshall1998}
{Marshall} F.~E.,  {Gotthelf} E.~V.,  {Zhang} W.,  {Middleditch} J.,   {Wang}
  Q.~D.,  1998, \mn@doi [apjl] {10.1086/311381}, \href
  {https://ui.adsabs.harvard.edu/abs/1998ApJ...499L.179M} {499, L179}

\bibitem[\protect\citeauthoryear{{Marshall}, {Guillemot}, {Kust Harding},
  {Martin}  \& {Smith}}{{Marshall} et~al.}{2016}]{Marshall2016}
{Marshall} F.~E.,  {Guillemot} L.,  {Kust Harding} A.,  {Martin} P.,   {Smith}
  D.~A.,  2016, in American Astronomical Society Meeting Abstracts \#227. p.
  423.04

\bibitem[\protect\citeauthoryear{{Matteucci}}{{Matteucci}}{2014}]{Matteucci2014}
{Matteucci} F.,  2014, in {Bland-Hawthorn} J.,  {Freeman} K.,   {Matteucci} F.,
   eds,  Saas-Fee Advanced Course Vol. 37, Saas-Fee Advanced Course. p.~145
  (\mn@eprint {arXiv} {0804.1492}), \mn@doi{10.1007/978-3-642-41720-7_2}

\bibitem[\protect\citeauthoryear{{McConnell}, {McCulloch}, {Hamilton}, {Ables},
  {Hall}, {Jacka}  \& {Hunt}}{{McConnell} et~al.}{1991}]{McConnell1991}
{McConnell} D.,  {McCulloch} P.~M.,  {Hamilton} P.~A.,  {Ables} J.~G.,  {Hall}
  P.~J.,  {Jacka} C.~E.,   {Hunt} A.~J.,  1991, \mn@doi [mnras]
  {10.1093/mnras/249.4.654}, \href
  {https://ui.adsabs.harvard.edu/abs/1991MNRAS.249..654M} {249, 654}

\bibitem[\protect\citeauthoryear{{McCulloch}, {Hamilton}, {Ables}  \&
  {Hunt}}{{McCulloch} et~al.}{1983}]{McCulloch1983}
{McCulloch} P.~M.,  {Hamilton} P.~A.,  {Ables} J.~G.,   {Hunt} A.~J.,  1983,
  \mn@doi [nat] {10.1038/303307a0}, \href
  {https://ui.adsabs.harvard.edu/abs/1983Natur.303..307M} {303, 307}

\bibitem[\protect\citeauthoryear{{Men}, {Barr}, {Clark}, {Carli}  \&
  {Desvignes}}{{Men} et~al.}{2023}]{men2023}
{Men} Y.,  {Barr} E.,  {Clark} C.~J.,  {Carli} E.,   {Desvignes} G.,  2023,
  \mn@doi [\aap] {10.1051/0004-6361/202347356}, \href
  {https://ui.adsabs.harvard.edu/abs/2023A&A...679A..20M} {679, A20}

\bibitem[\protect\citeauthoryear{{Middleditch} \& {Pennypacker}}{{Middleditch}
  \& {Pennypacker}}{1985}]{MiddleditchPennypacker1985}
{Middleditch} J.,  {Pennypacker} C.,  1985, \mn@doi [\nat] {10.1038/313659a0},
  \href {https://ui.adsabs.harvard.edu/abs/1985Natur.313..659M} {313, 659}

\bibitem[\protect\citeauthoryear{{Morello} et~al.,}{{Morello}
  et~al.}{2019}]{Morello2019}
{Morello} V.,  et~al., 2019, \mn@doi [\mnras] {10.1093/mnras/sty3328}, \href
  {https://ui.adsabs.harvard.edu/abs/2019MNRAS.483.3673M} {483, 3673}

\bibitem[\protect\citeauthoryear{{Morello}, {Barr}, {Stappers}, {Keane}  \&
  {Lyne}}{{Morello} et~al.}{2020}]{Morello2020}
{Morello} V.,  {Barr} E.~D.,  {Stappers} B.~W.,  {Keane} E.~F.,   {Lyne} A.~G.,
   2020, \mn@doi [\mnras] {10.1093/mnras/staa2291}, \href
  {https://ui.adsabs.harvard.edu/abs/2020MNRAS.497.4654M} {497, 4654}

\bibitem[\protect\citeauthoryear{{Padmanabh} et~al.,}{{Padmanabh}
  et~al.}{2023}]{Padmanabh2023}
{Padmanabh} P.~V.,  et~al., 2023, \mn@doi [\mnras] {10.1093/mnras/stad1900},
  \href {https://ui.adsabs.harvard.edu/abs/2023MNRAS.524.1291P} {524, 1291}

\bibitem[\protect\citeauthoryear{{Pakmor} et~al.,}{{Pakmor}
  et~al.}{2022}]{Pakmor2022}
{Pakmor} R.,  et~al., 2022, \mn@doi [\mnras] {10.1093/mnras/stac717}, \href
  {https://ui.adsabs.harvard.edu/abs/2022MNRAS.512.3602P} {512, 3602}

\bibitem[\protect\citeauthoryear{{Pan} et~al.,}{{Pan} et~al.}{2021}]{Pan2021}
{Pan} Z.,  et~al., 2021, \mn@doi [apjl] {10.3847/2041-8213/ac0bbd}, \href
  {https://ui.adsabs.harvard.edu/abs/2021ApJ...915L..28P} {915, L28}

\bibitem[\protect\citeauthoryear{{Papitto} et~al.,}{{Papitto}
  et~al.}{2013}]{Papitto2013}
{Papitto} A.,  et~al., 2013, \mn@doi [\nat] {10.1038/nature12470}, \href
  {https://ui.adsabs.harvard.edu/abs/2013Natur.501..517P} {501, 517}

\bibitem[\protect\citeauthoryear{{Pennock} et~al.,}{{Pennock}
  et~al.}{2021}]{Pennock2021}
{Pennock} C.~M.,  et~al., 2021, \mn@doi [\mnras] {10.1093/mnras/stab1858},
  \href {https://ui.adsabs.harvard.edu/abs/2021MNRAS.506.3540P} {506, 3540}

\bibitem[\protect\citeauthoryear{{Piatti}, {Alfaro}  \&
  {Cantat-Gaudin}}{{Piatti} et~al.}{2019}]{Piatti2019}
{Piatti} A.~E.,  {Alfaro} E.~J.,   {Cantat-Gaudin} T.,  2019, \mn@doi [\mnras]
  {10.1093/mnrasl/sly240}, \href
  {https://ui.adsabs.harvard.edu/abs/2019MNRAS.484L..19P} {484, L19}

\bibitem[\protect\citeauthoryear{{Price}}{{Price}}{2016}]{Price2016}
{Price} D.~C.,  2016, {PyGSM: Python interface to the Global Sky Model},
  Astrophysics Source Code Library, record ascl:1603.013 (\mn@eprint {ascl}
  {1603.013})

\bibitem[\protect\citeauthoryear{{Radhakrishnan} \&
  {Srinivasan}}{{Radhakrishnan} \& {Srinivasan}}{1982}]{Radhakrishnan1982}
{Radhakrishnan} V.,  {Srinivasan} G.,  1982, Current Science, \href
  {https://ui.adsabs.harvard.edu/abs/1982CSci...51.1096R} {51, 1096}

\bibitem[\protect\citeauthoryear{{Ransom}}{{Ransom}}{2011}]{Ransom2011}
{Ransom} S.,  2011, {PRESTO: PulsaR Exploration and Search TOolkit},
  Astrophysics Source Code Library, record ascl:1107.017 (\mn@eprint {ascl}
  {1107.017})

\bibitem[\protect\citeauthoryear{{Ransom}, {Hessels}, {Stairs}, {Freire},
  {Camilo}, {Kaspi}  \& {Kaplan}}{{Ransom} et~al.}{2005}]{Ransom2005}
{Ransom} S.~M.,  {Hessels} J. W.~T.,  {Stairs} I.~H.,  {Freire} P. C.~C.,
  {Camilo} F.,  {Kaspi} V.~M.,   {Kaplan} D.~L.,  2005, \mn@doi [Science]
  {10.1126/science.1108632}, \href
  {https://ui.adsabs.harvard.edu/abs/2005Sci...307..892R} {307, 892}

\bibitem[\protect\citeauthoryear{{Reardon} et~al.,}{{Reardon}
  et~al.}{2023}]{Reardon2023}
{Reardon} D.~J.,  et~al., 2023, \mn@doi [\apjl] {10.3847/2041-8213/acdd02},
  \href {https://ui.adsabs.harvard.edu/abs/2023ApJ...951L...6R} {951, L6}

\bibitem[\protect\citeauthoryear{{Reyes-Iturbide}, {Rosado}  \&
  {Vel{\'a}zquez}}{{Reyes-Iturbide} et~al.}{2008}]{Reyes2008}
{Reyes-Iturbide} J.,  {Rosado} M.,   {Vel{\'a}zquez} P.~F.,  2008, \mn@doi
  [\aj] {10.1088/0004-6256/136/5/2011}, \href
  {https://ui.adsabs.harvard.edu/abs/2008AJ....136.2011R} {136, 2011}

\bibitem[\protect\citeauthoryear{{Ridley}, {Crawford}, {Lorimer}, {Bailey},
  {Madden}, {Anella}  \& {Chennamangalam}}{{Ridley} et~al.}{2013}]{Ridley2013}
{Ridley} J.~P.,  {Crawford} F.,  {Lorimer} D.~R.,  {Bailey} S.~R.,  {Madden}
  J.~H.,  {Anella} R.,   {Chennamangalam} J.,  2013, \mn@doi [mnras]
  {10.1093/mnras/stt709}, \href
  {https://ui.adsabs.harvard.edu/abs/2013MNRAS.433..138R} {433, 138}

\bibitem[\protect\citeauthoryear{{Ridolfi} et~al.,}{{Ridolfi}
  et~al.}{2021}]{Ridolfi2021}
{Ridolfi} A.,  et~al., 2021, \mn@doi [mnras] {10.1093/mnras/stab790}, \href
  {https://ui.adsabs.harvard.edu/abs/2021MNRAS.504.1407R} {504, 1407}

\bibitem[\protect\citeauthoryear{{Ridolfi} et~al.,}{{Ridolfi}
  et~al.}{2022}]{Ridolfi2022}
{Ridolfi} A.,  et~al., 2022, \mn@doi [\aap] {10.1051/0004-6361/202143006},
  \href {https://ui.adsabs.harvard.edu/abs/2022A&A...664A..27R} {664, A27}

\bibitem[\protect\citeauthoryear{{Robitaille} \& {Bressert}}{{Robitaille} \&
  {Bressert}}{2012}]{Robitaille2012}
{Robitaille} T.,  {Bressert} E.,  2012, {APLpy: Astronomical Plotting Library
  in Python}, Astrophysics Source Code Library, record ascl:1208.017

\bibitem[\protect\citeauthoryear{{Rosado}, {Laval}, {Le Coarer}, {Boulesteix},
  {Georgelin}  \& {Marcelin}}{{Rosado} et~al.}{1993}]{Rosado1993}
{Rosado} M.,  {Laval} A.,  {Le Coarer} E.,  {Boulesteix} J.,  {Georgelin}
  Y.~P.,   {Marcelin} M.,  1993, \aap, \href
  {https://ui.adsabs.harvard.edu/abs/1993A&A...272..541R} {272, 541}

\bibitem[\protect\citeauthoryear{{Sasaki} et~al.,}{{Sasaki}
  et~al.}{2022}]{Sasaki2022}
{Sasaki} M.,  et~al., 2022, \mn@doi [\aap] {10.1051/0004-6361/202141054}, \href
  {https://ui.adsabs.harvard.edu/abs/2022A&A...661A..37S} {661, A37}

\bibitem[\protect\citeauthoryear{{Scargle}}{{Scargle}}{1969}]{Scargle1969}
{Scargle} J.~D.,  1969, \mn@doi [apj] {10.1086/149978}, \href
  {https://ui.adsabs.harvard.edu/abs/1969ApJ...156..401S} {156, 401}

\bibitem[\protect\citeauthoryear{{Seward}, {Harnden}  \& {Helfand}}{{Seward}
  et~al.}{1984}]{Seward1984}
{Seward} F.~D.,  {Harnden} F.~R. J.,   {Helfand} D.~J.,  1984, \mn@doi [apjl]
  {10.1086/184388}, \href
  {https://ui.adsabs.harvard.edu/abs/1984ApJ...287L..19S} {287, L19}

\bibitem[\protect\citeauthoryear{{Stappers} \& {Kramer}}{{Stappers} \&
  {Kramer}}{2016}]{Stappers2016}
{Stappers} B.,  {Kramer} M.,  2016, in MeerKAT Science: On the Pathway to the
  SKA. p.~9, \mn@doi{10.22323/1.277.0009}

\bibitem[\protect\citeauthoryear{{Staveley-Smith} et~al.,}{{Staveley-Smith}
  et~al.}{1996}]{Staveley-Smith1996}
{Staveley-Smith} L.,  et~al., 1996, \mn@doi [\pasa]
  {10.1017/S1323358000020919}, \href
  {https://ui.adsabs.harvard.edu/abs/1996PASA...13..243S} {13, 243}

\bibitem[\protect\citeauthoryear{{Tauris} \& {van den Heuvel}}{{Tauris} \& {van
  den Heuvel}}{2006}]{Tauris2006}
{Tauris} T.~M.,  {van den Heuvel} E.~P.~J.,  2006, in , Vol.~39, Compact
  stellar X-ray sources.
Cambridge University Press, pp 623--665,
  \mn@doi{10.48550/arXiv.astro-ph/0303456}

\bibitem[\protect\citeauthoryear{{Verbunt}, {Igoshev}  \& {Cator}}{{Verbunt}
  et~al.}{2017}]{Verbunt2017}
{Verbunt} F.,  {Igoshev} A.,   {Cator} E.,  2017, \mn@doi [\aap]
  {10.1051/0004-6361/201731518}, \href
  {https://ui.adsabs.harvard.edu/abs/2017A&A...608A..57V} {608, A57}

\bibitem[\protect\citeauthoryear{{Wang}, {Gotthelf}, {Chu}  \& {Dickel}}{{Wang}
  et~al.}{2001}]{Wang2001}
{Wang} Q.~D.,  {Gotthelf} E.~V.,  {Chu} Y.~H.,   {Dickel} J.~R.,  2001, \mn@doi
  [\apj] {10.1086/322392}, \href
  {https://ui.adsabs.harvard.edu/abs/2001ApJ...559..275W} {559, 275}

\bibitem[\protect\citeauthoryear{{Wang} et~al.,}{{Wang}
  et~al.}{2022}]{Wang2022}
{Wang} Y.,  et~al., 2022, \mn@doi [apj] {10.3847/1538-4357/ac61dc}, \href
  {https://ui.adsabs.harvard.edu/abs/2022ApJ...930...38W} {930, 38}

\bibitem[\protect\citeauthoryear{{Weltevrede}}{{Weltevrede}}{2016}]{Weltevrede2016}
{Weltevrede} P.,  2016, \mn@doi [\aap] {10.1051/0004-6361/201527950}, \href
  {https://ui.adsabs.harvard.edu/abs/2016A&A...590A.109W} {590, A109}

\bibitem[\protect\citeauthoryear{{Wenger} et~al.,}{{Wenger}
  et~al.}{2000}]{Wenger2000}
{Wenger} M.,  et~al., 2000, \mn@doi [\aaps] {10.1051/aas:2000332}, \href
  {https://ui.adsabs.harvard.edu/abs/2000A&AS..143....9W} {143, 9}

\bibitem[\protect\citeauthoryear{{Williams}, {Chu}, {Dickel}, {Gruendl},
  {Seward}, {Guerrero}  \& {Hobbs}}{{Williams} et~al.}{2005}]{Williams2005}
{Williams} R.~M.,  {Chu} Y.~H.,  {Dickel} J.~R.,  {Gruendl} R.~A.,  {Seward}
  F.~D.,  {Guerrero} M.~A.,   {Hobbs} G.,  2005, \mn@doi [\apj]
  {10.1086/431349}, \href
  {https://ui.adsabs.harvard.edu/abs/2005ApJ...628..704W} {628, 704}

\bibitem[\protect\citeauthoryear{{Xie} et~al.,}{{Xie} et~al.}{2019}]{Xie2019}
{Xie} Y.-W.,  et~al., 2019, \mn@doi [Research in Astronomy and Astrophysics]
  {10.1088/1674-4527/19/7/103}, \href
  {https://ui.adsabs.harvard.edu/abs/2019RAA....19..103X} {19, 103}

\bibitem[\protect\citeauthoryear{{Xu} et~al.,}{{Xu} et~al.}{2023}]{Xu2023}
{Xu} H.,  et~al., 2023, \mn@doi [Research in Astronomy and Astrophysics]
  {10.1088/1674-4527/acdfa5}, \href
  {https://ui.adsabs.harvard.edu/abs/2023RAA....23g5024X} {23, 075024}

\bibitem[\protect\citeauthoryear{{Yao}, {Manchester}  \& {Wang}}{{Yao}
  et~al.}{2017}]{Yao2017}
{Yao} J.~M.,  {Manchester} R.~N.,   {Wang} N.,  2017, \mn@doi [\apj]
  {10.3847/1538-4357/835/1/29}, \href
  {https://ui.adsabs.harvard.edu/abs/2017ApJ...835...29Y} {835, 29}

\bibitem[\protect\citeauthoryear{{Yew} et~al.,}{{Yew} et~al.}{2021}]{Yew2021}
{Yew} M.,  et~al., 2021, \mn@doi [\mnras] {10.1093/mnras/staa3382}, \href
  {https://ui.adsabs.harvard.edu/abs/2021MNRAS.500.2336Y} {500, 2336}

\bibitem[\protect\citeauthoryear{{Zanardo}, {Staveley-Smith}, {Gaensler},
  {Indebetouw}, {Ng}, {Matsuura}  \& {Tzioumis}}{{Zanardo}
  et~al.}{2018}]{Zanardo2018}
{Zanardo} G.,  {Staveley-Smith} L.,  {Gaensler} B.~M.,  {Indebetouw} R.,  {Ng}
  C.~Y.,  {Matsuura} M.,   {Tzioumis} A.~K.,  2018, \mn@doi [\apjl]
  {10.3847/2041-8213/aacc2a}, \href
  {https://ui.adsabs.harvard.edu/abs/2018ApJ...861L...9Z} {861, L9}

\bibitem[\protect\citeauthoryear{{Zangrandi} et~al.,}{{Zangrandi}
  et~al.}{2024}]{Zangrandi2024}
{Zangrandi} F.,  et~al., 2024, \mn@doi [arXiv e-prints]
  {10.48550/arXiv.2401.17307}, \href
  {https://ui.adsabs.harvard.edu/abs/2024arXiv240117307Z} {p. arXiv:2401.17307}

\bibitem[\protect\citeauthoryear{{Zhang}, {Dai}, {Hobbs}, {Staveley-Smith},
  {Manchester}, {Russell}, {Zanardo}  \& {Wu}}{{Zhang}
  et~al.}{2018}]{Zhang2018}
{Zhang} S.~B.,  {Dai} S.,  {Hobbs} G.,  {Staveley-Smith} L.,  {Manchester}
  R.~N.,  {Russell} C.~J.,  {Zanardo} G.,   {Wu} X.~F.,  2018, \mn@doi [\mnras]
  {10.1093/mnras/sty1573}, \href
  {https://ui.adsabs.harvard.edu/abs/2018MNRAS.479.1836Z} {479, 1836}

\bibitem[\protect\citeauthoryear{{Zheng} et~al.,}{{Zheng}
  et~al.}{2017}]{Zheng2017}
{Zheng} H.,  et~al., 2017, \mn@doi [\mnras] {10.1093/mnras/stw2525}, \href
  {https://ui.adsabs.harvard.edu/abs/2017MNRAS.464.3486Z} {464, 3486}

\bibitem[\protect\citeauthoryear{{Zhu} et~al.,}{{Zhu} et~al.}{2014}]{Zhu2014}
{Zhu} W.~W.,  et~al., 2014, \mn@doi [\apj] {10.1088/0004-637X/781/2/117}, \href
  {https://ui.adsabs.harvard.edu/abs/2014ApJ...781..117Z} {781, 117}

\bibitem[\protect\citeauthoryear{{van Heerden}, {Karastergiou}  \&
  {Roberts}}{{van Heerden} et~al.}{2017}]{vanHeerden2017}
{van Heerden} E.,  {Karastergiou} A.,   {Roberts} S.~J.,  2017, \mn@doi
  [\mnras] {10.1093/mnras/stw3068}, \href
  {https://ui.adsabs.harvard.edu/abs/2017MNRAS.467.1661V} {467, 1661}

\bibitem[\protect\citeauthoryear{{van Straten}, {Demorest}  \& {Oslowski}}{{van
  Straten} et~al.}{2012}]{VanStraten2012}
{van Straten} W.,  {Demorest} P.,   {Oslowski} S.,  2012, \mn@doi [Astronomical
  Research and Technology] {10.48550/arXiv.1205.6276}, \href
  {https://ui.adsabs.harvard.edu/abs/2012AR&T....9..237V} {9, 237}

\bibitem[\protect\citeauthoryear{{van den Eijnden} \& {Rajwade}}{{van den
  Eijnden} \& {Rajwade}}{2024}]{Eijnden2024}
{van den Eijnden} J.,  {Rajwade} K.,  2024, \mn@doi [Research Notes of the
  American Astronomical Society] {10.3847/2515-5172/ad22d2}, \href
  {https://ui.adsabs.harvard.edu/abs/2024RNAAS...8...34V} {8, 34}

\makeatother
\end{thebibliography}

\bsp	
\label{lastpage}
\end{document}